 \documentclass[sigconf,screen]{acmart}

\AtBeginDocument{%
  \providecommand\BibTeX{{%
    Bib\TeX}}}

\copyrightyear{2026}
\acmYear{2026}
\setcopyright{cc}
\setcctype{by}
\acmConference[ASE '26]{Proceedings of the 41st IEEE/ACM International Conference on Automated Software Engineering}{October 12--16, 2026}{Munich, Germany}
\acmBooktitle{Proceedings of the 41st IEEE/ACM International Conference on Automated Software Engineering (ASE '26), October 12--16, 2026, Munich, Germany}
\acmDOI{10.1145/3832783.3837529}
\acmISBN{979-8-4007-2882-2/2026/10}

\usepackage{xcolor}
\usepackage{algorithmic}
\usepackage[linesnumbered,ruled,vlined]{algorithm2e}
\usepackage{url}
\usepackage{balance}
\usepackage{xfrac}
\usepackage{subcaption}
\usepackage{textcomp}
\usepackage{xcolor}
\def\BibTeX{{\rm B\kern-.05em{\sc i\kern-.025em b}\kern-.08em
    T\kern-.1667em\lower.7ex\hbox{E}\kern-.125emX}}
 \usepackage{booktabs}
 \usepackage{multirow}

\usepackage[utf8]{inputenc}%added by XZ to have accent in the names like Barré ..
\usepackage{graphicx}
\usepackage{tabularx}
\usepackage{textcomp}
\usepackage{xcolor}
\usepackage{enumerate}
\usepackage{enumitem} %for enumerate "wide" option
\usepackage{wrapfig}
\usepackage{tcolorbox}

\usepackage{times}
\usepackage{soul}
\usepackage{url}
\usepackage{amsmath,amsthm,amsfonts}
\usepackage{caption}
\usepackage{booktabs}
\usepackage[absolute,overlay]{textpos}% by XZ: for the text in the top margin..

\newcount\Comments  % 0 suppresses notes to selves in text
\usepackage{color}
\definecolor{darkgreen}{rgb}{0,0.5,0}
\definecolor{purple}{rgb}{1,0,1}
\newcommand{\kibitz}[2]{\ifnum\Comments=1\textcolor{#1}{#2}\fi}
\newcommand{\approach}{EPIK}
\newtheorem{definition}{Definition}
\newtheorem{example}{Example}

\usepackage{tikz}

\AtBeginDocument{%
  \providecommand\BibTeX{{%
    \normalfont B\kern-0.5em{\scshape i\kern-0.25em b}\kern-0.8em\TeX}}}

\newcommand{\squishlist}{
   \begin{list}{$\bullet$}
    { \setlength{\itemsep}{2pt}%2
    \setlength{\parsep}{0pt}
      \setlength{\topsep}{5pt}     \setlength{\partopsep}{0pt}
      \setlength{\leftmargin}{1.35em} \setlength{\labelwidth}{1em}
      \setlength{\labelsep}{0.5em} } }

\newcommand{\squishlisttwo}{
   \begin{list}{$\bullet$}
    { \setlength{\itemsep}{0pt}    \setlength{\parsep}{0pt}
      \setlength{\topsep}{0pt}     \setlength{\partopsep}{0pt}
      \setlength{\leftmargin}{1.35em} \setlength{\labelwidth}{1em}
      \setlength{\labelsep}{0.5em} } }

\newcommand{\squishend}{
    \end{list}  }

\newcommand{\changed}[1]{\textcolor{black}{#1}}

\begin{document}

% \title[Learning Model Parameters from System-Level Observations for Quantitative Verification]{
\title[Leveraging System-Level Observations to Learn Model Parameters for Quantitative Verification]
{Leveraging System-Level Observations to Inform Bayesian Learning of Model Parameters for Quantitative Verification}

\author{Simos Gerasimou}
\affiliation{%
  \institution{Cyprus University of Technology, Cyprus}
  \city{}
  \country{}}
  \email{simos.gerasimou@cut.ac.cy}

\author{Xingyu~Zhao}
\affiliation{%
  \institution{Wuhan University, China}
  \city{}
  \country{}}
  \email{xingyu.zhao@whu.edu.cn}



\begin{abstract}
% Combining Bayesian learning and quantitative verification is a powerful toolset for analysing key quantitative properties of software systems, like reliability and response time. However, the accuracy and robustness of verification results strongly depend on the prior knowledge (PK) underlying Bayesian inference. 
% This knowledge reflects original beliefs about the probability of events and typically depends on domain expertise. 
% Using inaccurate or uninformative PK can negatively affect quantitative analysis, yielding incorrect verification results. 
% Our \approach\ approach tackles this important challenge by eliciting and embedding PK in %the framework of 
% quantitative verification equipped with Bayesian estimators. 
% Unlike %state-of-the-art 
% existing approaches that require PK on formal model transition parameters, \approach\ leverages system-level properties that are directly observable and are linked to real-world semantics. 
% % Then, 
% \approach\ formulates a \changed{twofold} optimisation problem to derive the distributions of unknown transition parameters and then embeds these distributions to verify new or difficult-to-measure (elusive) properties. 
% The detailed experimental evaluation using multiple variants of real-world case studies and diverse \approach\ instantiations shows
% % \approach's 
% its effectiveness, flexibility and generality.
Combining Bayesian learning and quantitative verification is a powerful toolset for analysing key quantitative properties of software systems, like reliability and response time. However, the accuracy and robustness of verification results strongly depend on the prior knowledge (PK) underlying Bayesian inference. This knowledge reflects original beliefs about the probability of events and typically depends on domain expertise. Using inaccurate or uninformative PK can negatively affect quantitative analysis, yielding incorrect verification results. Our \approach\ approach tackles this important challenge by eliciting and embedding PK in quantitative verification equipped with Bayesian estimators. Unlike existing approaches that require PK on formal model transition parameters, \approach\ leverages system-level properties that are directly observable and are linked to real-world semantics. \approach\ formulates a \changed{twofold} optimisation problem to derive the distributions of unknown transition parameters and then embeds these distributions to verify new or difficult-to-measure (elusive) properties. The detailed experimental evaluation using multiple variants of real-world case studies and diverse \approach\ instantiations shows its effectiveness, flexibility and generality.
\end{abstract}

% \begin{IEEEkeywords}
% parametric model checking; software performability; nonfunctional software properties; Markov models
% \end{IEEEkeywords}}

% Add keywords here (comma-separated):

\maketitle

\section{Introduction}
\label{sec_intro}

Modern software-controlled systems, such as robotics, environmental and healthcare monitoring applications, and cloud-based services, operate in complex and uncertain environments characterised by workloads, operational profiles, failures and resource availability that are stochastic in nature~\cite{de2013software,weyns2023towards}.
Designing and analysing the performance, dependability and other nonfunctional properties of such systems when deployed in these uncertain or, potentially, adversarial environments can be facilitated by rigorous model-based verification methods~\cite{kwiatkowska2004modelling}.
More specifically, probabilistic models, including Markov chains~\cite{gerasimou_search-based_2015,kwiatkowska2007stochastic}, queueing networks~\cite{cortellessa2011model,balsamo2003review} and stochastic Petri nets~\cite{perez2010performance,tigane2022quantitative} can be leveraged to model and analyse the behaviour and operating environment of such systems. 

This paper focuses on probabilistic model checking (PMC), a formal method for automatically verifying quantitative aspects of stochastic systems~\cite{baier2008principles}. 
PMC involves the construction of a Markov chain model that encodes the system's behaviour over time, including states the system can reside in, the possible transitions between these states and information about the likelihood (probability) or timing (rate) of these transitions. 
Given properties of the required behaviour of these systems (e.g., reliability, performance) formally specified in temporal logic, the systematic analysis of the system model through automated algorithms enables assessing if the properties are satisfied~\cite{baier2008principles}.
Using mathematical reasoning to derive guarantees for achieving precisely defined levels of performance or efficiency is highly valuable for the design, analysis and adaptation of software-intensive cyber-physical systems~\cite{kwiatkowska_probabilistic_2022,calinescu2017synthesis,calinescu2017using}.

Recent advances improve the PMC efficiency and scalability~\cite{filieri_formal_2012,fang2021fast,jansen_accelerating_2014,evangelidis2026accelerating}, enabling %its employment for 
the analysis of more complex models and properties of software product lines~\cite{ghezzi2013model}, software architectures~\cite{moreno2014impact} and cyber-physical systems~\cite{zhao_towards_2019}. 
As a verification method, however, the verification results strongly depend on the veracity of the Markov model encoding the behaviour of the target system and, especially, the parameters (probability/rate) capturing the transitions between model states~\cite{kwiatkowska2007stochastic}. 
When these model parameters accurately reflect the current system behaviour, the verification results faithfully represent quantitative system aspects. 
In contrast, inaccurate or outdated model parameter values yield misleading results that unavoidably lead to incorrect engineering and adaptation decisions~\cite{calinescu_self_adaptive_2012}. 
% quantify  provide formal guarantees about to 

Approaches to improve analysis and PMC reasoning use Bayesian methods to incorporate 
prior (domain) knowledge at design time~\cite{jha2009bayesian} and update the Markov model using system observations at runtime~\cite{epifani_model_2009,filieri_supporting_2016,calinescu_adaptive_2014}.
Notwithstanding their benefits, a fundamental premise underpinning these Bayesian-based approaches is that prior knowledge about the transition probability/rate between two model states can be extracted from domain experts or %lab simulations and 
past similar system executions~\cite{epifani_model_2009}. 
While this may be desirable, demanding domain experts to define their knowledge about model transition parameters (e.g., the probability of a fruit-picking robot transitioning between the positioning and fruit picking states)
%setting itself to the appropriate position to pick the next fruit) 
is a challenging and non-trivial problem that needs fine-grained system knowledge~\cite{fang2022presto}. 
Providing biased or uninformative values for model transition parameters yields inaccurate analysis results, leading to design decisions that can be detrimental to the system's reliability and performance~\cite{bishop_toward_2011}. 

% We postulate that instead of defining these \textit{latent parameters} of the underlying formal model, domain experts typically possess rich \textit{system-wide knowledge} about properties of system behaviour~\cite{GUINDON1990279}.
Instead of enforcing the definition of values for transition parameters of the underlying formal model, we \changed{propose employing} the rich \textit{system-wide knowledge} about properties of system behaviour typically possessed by domain experts~\cite{GUINDON1990279}. 
These properties are directly observable and correspond to intuitive real-world semantics and actions.
This can be expressed, for instance, as the reliability of the fruit-picking robot completing its task successfully~\cite{fang2022presto} or the expected execution time of a service-based system that uses cloud-based services~\cite{gerasimou_search-based_2015}. 
Since such system-wide knowledge embodies real-world semantics and is directly observable from past system executions, it aligns more naturally with the available domain expertise~\cite{xie2021survey,li2020explanations}.
Hence, this system-wide knowledge can be analysed to extract appropriate values for the model transition parameters. 

% \begin{figure}[t]
%	 	% \vspace*{-5mm}
%	 	\centering\includegraphics[width=\linewidth]{images/EPIKapproach.png}
%	 	\vspace*{-3mm}
%	 	\caption{EPIK overview showing its main two steps for knowledge elicitation and knowledge embedding}
%	 	\label{fig:approach}
%	 \end{figure}

Driven by this insight, we introduce \approach, a Bayesian-based PMC approach that leverages expert knowledge of system-wide properties (termed \emph{PK-informed}) for the systematic elicitation and embedding of model transition parameters. 
% \approach, whose high-level overview is shown in Fig.~\ref{fig:epik}, comprises two main steps.
\approach\ comprises the following key stages.
First, during the \emph{knowledge elicitation} stage, prior knowledge from domain experts, corresponding to past observations of system-wide properties (e.g., reliability, response time), each expressed as a probability distribution, initiates a multi-objective search-based elicitation problem~\cite{coello2007evolutionary}.
The synthesised Pareto set signifies model transition parameters that best approximate the probability distributions of the system-wide PK-informed properties. %specified by domain experts. 
Then, during the \emph{knowledge embedding} stage, \approach\ exploits the synthesised Pareto set of transition parameters for the verification of \emph{elusive} system properties, i.e., system-wide properties that are novel or rare, or those for which gathering information is risky or expensive.
% Decision-makers can leverage \approach\ both to analyse these elusive properties and to inform the system design and implementation~\cite{filieri_probabilistic_2013,zhao2023bayesian} and for self-adaptation~\cite{calinescu_self_adaptive_2012}.

We evaluate \approach\ on multiple variants from two software-contro-lled systems from different application domains: 
(1) a fruit-picking robot (FPR)~\cite{fang2022presto} and 
(2) a service-based system from foreign exchange trading~\cite{gerasimou2018synthesis}, demonstrating the accuracy and effectiveness of \approach\ in extracting and embedding prior knowledge. 
To the best of our knowledge, \approach\ is the first tool-supported solution that considers the problem of extracting knowledge for the unknown transition parameters of a probabilistic model from system-wide properties.

The real-world applicability of \approach\ follows the broad adoption of similar work applying Bayesian learning in probabilistic analysis~\cite{zhao2023bayesian,jha2009bayesian,epifani_model_2009,filieri_supporting_2016}, and the various models of software systems available in the repositories of probabilistic model checkers~\cite{kwiatkowska2011prism,dehnert2017storm}.
\changed{We also emphasise \approach's positioning with respect to downstream Bayesian learning approaches.
\approach\ is not a posterior inference approach that uses runtime observations.
Instead, \approach\ operates entirely during design-time (preparation phase) to systematically synthesise mathematically grounded prior distributions from observable, high-level system properties. 
The elicited distributions produced by \approach\ serve as the initial priors ($PK$) for downstream Bayesian estimators (e.g., KAMI~\cite{epifani_model_2009}). 
This pipeline prevents the biased or inaccurate initialisations that frequently cause subsequent Bayesian approaches to fail or produce distorted verification results.
}

The main contributions of our paper are:
\\ \noindent $\bullet$ 
The \approach\ approach to elicit and embed prior knowledge for Bayesian-based PMC that leverages \emph{PK-informed} system-wide properties and enables the verification of \emph{elusive} properties;
\\ \noindent $\bullet$ 
An extensive \approach\ evaluation on several variants of two real-world problems, for a wide variety of PK-informed properties and a set of unknown transition parameters. 
\\ \noindent $\bullet$ 
A prototype open-source \approach\ tool and case study repository, available at 
\url{https://github.com/gerasimou/EPIK}.
% \url{https://anonymous.4open.science/r/EPIK-BAD4}.

The remainder of this paper is structured as follows: Section~\ref{sec_preliminaries} presents the required background on PMC and Bayesian learning.
Section~\ref{sec_example} introduces a running example to illustrate \approach, which is detailed in Section~\ref{sec_method}. 
Section~\ref{sec_eva} describes \approach's implementation and evaluation.
Section~\ref{sec_related_work} discusses related work, and Section~\ref{sec_conclusion} summarises our results and
suggests future research directions.

\section{Preliminaries}
\label{sec_preliminaries}

% \subsection{Probabilistic Model Checking}
% \label{sec_preliminaries_pmc}

\noindent 
\textbf{Probabilistic Model Checking.}
Probabilistic model checking (PMC) is a formal method for assessing quantitative properties, e.g., reliability, performance, cost, of systems exhibiting stochastic behaviour~\cite{baier2008principles}. 
Markov chains capture the system's stochastic behaviour, enabling the analysis of properties encoded as formal logic specifications. 
A Markov chain is a tuple $M=(S,s_0,\delta, L)$, where $S$ is a finite set of states, $s_0\in S$ is the initial state, $\delta$ is the state-transition function:
\squishlisttwo
\item $\delta : S\times S\rightarrow [0,1]$ for discrete-time Markov chains (DTMCs), with $\delta(s_i,s_j)=p_{ij}$ giving the transition probability between states $s_i,s_j\in S$, and $\sum_{s_j\in S} \delta(s_i,s_j)=1$;
\item $\delta : S \times S \rightarrow \mathbb{R}_{\geq 0}$ for continuous-time Markov chains (CTMCs); $\delta(s_i,s_j)=r_{ij}$ gives the rate of transition between states $s_i,s_j\in S$.
\squishend
and $L:S\rightarrow 2^{AP}$ is a labelling function assigning to each state a set of atomic propositions from  $AP$.

PMC supports both DTMCs and CTMCs as modelling formalisms. 
DTMCs are widely used to represent systems with discrete, sequential behaviour, where transitions occur at discrete time steps. 
CTMCs model systems with continuous, time-dependent behaviour, where transitions can occur at any point in time. 
%PMC extends its capabilities to analyse properties in CTMCs by considering transition rates between states and steady-state probabilities. 
PMC can analyse properties in DTMCs/CTMCs by considering the probabilities of reaching safe/unsafe states and computing expected values (cost/rewards). 
%By supporting both DTMCs and CTMCs, PMC offers a comprehensive framework for analysing a wide range of systems with stochastic behaviour, accommodating both discrete and continuous aspects. 
%This flexibility allows for the verification and evaluation of system properties in diverse domains.

Markov model states are labelled with atomic propositions that hold in those states. The properties to verify are expressed in temporal logic over these atomic propositions, e.g., probabilistic temporal tree logic (PCTL)~\cite{bianco_alfaro_1995,hansson1994logic} for DTMCs and continuous stochastic logic (CSL)~\cite{aziz1996verifying} for CTMCs. 
State-of-the-art probabilistic model checkers (PRISM~\cite{kwiatkowska2011prism}, Storm~\cite{dehnert2017storm}) implement efficient PMC algorithms for verifying  Markov chain models and properties~\cite{kwiatkowska2007stochastic}.

\vspace{1mm}\noindent
\textbf{Parametric Model Checking.}\label{paraMC}
PMC based on DTMCs/CTMCs assumes that transition probabilities/rates are known constants, or can be estimated from existing data and experts at design time. 
% Given the complex and dynamic nature of modern systems, 
% \textit{Parametric} model checking  (ParaMC)~\cite{daws_symbolic_2005}  provides an efficient solution to establish more accurate estimations on the transition probabilities/rates through runtime monitoring using up-to-date data~\cite{filieri_run_time_2011,calinescu_self_adaptive_2012}.
%
\textit{Parametric} model checking  (ParaMC)~\cite{daws_symbolic_2005} can analyse DTMCs/CTMCs whose transition probabilities/rates are specified as functions over a set of parameters, termed \textit{parametric} DTMCs/CTMCs. 
The property under verification is given as a \textit{closed-form rational function} of these parameters~\cite{daws_symbolic_2005}. 
This formulation brings a practical advantage by dividing the verification process into two steps. 
The computationally intensive symbolic analysis can be performed offline without strict computational constraints~\cite{jansen_accelerating_2014}. 
Then, the closed-form symbolic expressions can be reused later, e.g., when new knowledge or data of the parameters is learned, without rerunning the PMC every time. 
%Then, at runtime, simple substitutions are required to replace the parameters in the closed-form expressions with the actual values learned. 
\approach\  leverages the closed-form expressions to efficiently derive PK for the unknown model transition parameters. 
%This is done by eliciting prior knowledge from experts regarding certain properties that can be confidently expressed based on historical data and the experts' knowledge.

\vspace{1mm}\noindent
\textbf{Bayesian Learning in Probabilistic Model Checking. }
Given the DTMC state $s_i$, the transition to the next state follows a \textit{categorical distribution}~\cite{epifani_model_2009}.
Due to the Markov property, the choice of the next state only depends on the current state, and the categorical distributions per state are \textit{independent}. 
% Hence, as we observe repeated transitions from state $i$, the repeated categorical process follows a \textit{multinomial distribution}. 
The categorical process of repeated transitions from state~$i$, follows a \textit{multinomial distribution}. 
% The Bayesian learning problem of transition parameters is reduced to the \textit{localised} learning of $k$ independent multinomial distributions, where $k$ is the total number of DTMC states~\cite{gelman_bayesian_2014,bernardo_bayesian_1994}. 
Thus, learning the transition parameters reduces to the \textit{localised} learning of $k$ independent multinomial distributions, where $k$ is the number of DTMC states~\cite{gelman_bayesian_2014}.

From a Bayesian inference perspective, the posterior estimation needs a statistical model (the \textit{likelihood} function) and \textit{a prior distribution}~\cite{berger1994overview}.
For the likelihood function, if we observe $n_{ij}$ transitions from state $s_i$ to state $s_j$ out of $n_i$ outgoing transitions from $s_i$ (%labelled as 
termed ``data'' in the equations below), the binomial likelihood is (omitting the combinatorial factor that cancels out in the Bayes formula):
\begin{equation}
	% \vspace{-0.5mm}
	\label{eq_binomial_likelihood}
	\mathit{Pr}(\textmd{data} \mid p_{ij}) = p_{ij}^{n_{ij}}(1-p_{ij})^{n_i-n_{ij}}
	% \vspace{-0.5mm}
\end{equation}
%where $p_{ij}=\delta(s_i,s_j)$.

% As usual, for mathematical convenience in Bayesian inference, the method in~\cite{epifani_model_2009} uses a \textit{conjugate} Beta prior distribution\footnote{Concretely, using a Dirichlet distribution as prior for the $i$-th transition matrix row, which is a multivariate generalisation of the Beta distribution.} for the likelihood. Specifically, it uses the \textit{canonical} parameterisation\footnote{A parameterisation where the shape parameters $\alpha$ and $\beta$ of the common $\mathit{Beta}(\alpha,\beta)$ are replaced by $n_i^{(0)}=\alpha+\beta$ and $p_{ij}^{(0)}=\alpha/(\alpha+\beta)$.} 

Typically, for mathematical convenience in Bayesian inference, a \textit{conjugate} Beta prior distribution\footnote{A Dirichlet distribution, which is a multivariate generalisation of the Beta distribution, can be used to model the prior of the $i$-th transition matrix row.} is used for the likelihood~\cite{epifani_model_2009}. Specifically, the \textit{canonical} parameterisation\footnote{The shape parameters $\alpha\!$, $\beta$ are replaced by $n_i^{(0)}\!=\!\alpha+\beta$, $p_{ij}^{(0)}=\alpha/(\alpha+\beta)$}
%\footnote{A parameterisation where the shape parameters $\alpha$ and $\beta$ of the common $\mathit{Beta}(\alpha,\beta)$ are replaced by $n_i^{(0)}=\alpha+\beta$ and $p_{ij}^{(0)}=\alpha/(\alpha+\beta)$.} 
%
\begin{equation}
	\label{eq_prior_beta}
	\mathit{Beta}\bigl(n_i^{(0)},p_{ij}^{(0)}\bigr)
	%\vspace{-1mm}
\end{equation}
of the Beta distribution allows an intuitive interpretation of $p_{ij}^{(0)}$ as the ``best prior probability guess'' and of $n_i^{(0)}$ as the sample size (strength) on which the prior estimation $p_{ij}^{(0)}$ is based on~\cite{epifani_model_2009,walter_imprecision_2009}.

After applying the Bayes rule and leveraging the conjugacy and canonical reparameterisation~\cite{gelman_bayesian_2014}, the posteriors are again a  $\mathit{Beta}\bigl(n_i^{(n_i)},p_{ij}^{(n_i)}\bigr)$ distribution with the updated parameters:
\vspace{-2mm}
\begin{align}
	n^{(n_i)}_i\!=\!n_i^{(0)}\!+\!n_i,\; \quad
	p_{ij}^{(n_i)}\!=\!\frac{n_i^{(0)}}{n_i^{(0)}+n_i}\!\cdot\! p_{ij}^{(0)}+\frac{n_i}{n_i^{(0)}+n_i}\!\cdot\!\frac{n_{ij}}{n_i}
	\label{eq_post_p_ij_fundamental}
	%\vspace{-2mm}
\end{align}
% where the superscript `$(0)$' reflects the fact that these values represent the knowledge before any observation is available. 
% In contrast, the superscript `${(n_i)}$' denotes the posterior parameters after observing $n_i$ outgoing transitions from state $s_i$. 
where the superscript `$(0)$' is the prior knowledge (before any observations), and the superscript `${(n_i)}$' gives the posterior parameters after observing $n_i$ outgoing transitions from state~$s_i$.

% Eq.~\eqref{eq_post_p_ij_fundamental} demonstrates that, upon observing $n_{ij}$ transitions out of $n_i$ total transitions, the posterior $p_{ij}^{(n_i)}$ can be expressed as a weighted sum of two terms: the prior estimate $p_{ij}^{(0)}$ and the new observations $\frac{n_{ij}}{n_i}$ (representing the frequency of transitions from $s_i$ to $s_j$). % in the data). 
Eq.~\eqref{eq_post_p_ij_fundamental} demonstrates that, upon observing $n_{ij}$ transitions out of $n_i$ total transitions, the posterior $p_{ij}^{(n_i)}$ is the weighted sum of the prior estimate $p_{ij}^{(0)}$ and the new observations $\frac{n_{ij}}{n_i}$ (encoding the frequency of transitions from $s_i$ to $s_j$). % in the data). 
The weights are proportional to $n_i^{(0)}$ 
% (the ``pseudo-count'' of the prior imaginary sample size) 
(the prior sample size -- strength) 
and $n_i$ (the ``actual count'' of data sample size). 
Smaller $n_i^{(0)}$ values indicate lower confidence in the priors, allowing the runtime data to influence the posteriors more. 
When $n_i^{(0)}\!\simeq\!0$, Eq.~\eqref{eq_post_p_ij_fundamental} simplifies to the Maximum Likelihood Estimation~\cite{epifani_model_2009}.
% Similarly, for CTMCs, where the Gamma-Poisson setup is typically applied
For CTMCs, where the Gamma-Poisson setup typically applies~\cite{bernardo_bayesian_1994},
%(cf. Proposition 5.4 and Example 5.5 in 
%\cite[pp.266--277]{bernardo_bayesian_1994})
Bayesian estimators for CTMC transition rates entail replacing the Beta priors with Gamma priors~\cite{filieri_formal_2012}.

\section{Motivating Example}\label{sec_example}
We illustrate \approach\ using an autonomous robot executing a fruit-picking task
% ~\cite{fang2022presto}.
adapted from~\cite{fang2022presto}.
% The robot is responsible for performing the following three operations in sequence:
The robot should perform the following operations: %in order:
(i) position itself %in the  location 
to collect the next fruit in its vicinity;
(ii) use its arm to collect the fruit; and 
(iii) when fruit collection is unsuccessful, decide whether to retry the collection by repositioning itself or abandon the task and move to the next fruit.

\begin{figure}[t]
	\centering\includegraphics[width=0.755\linewidth]{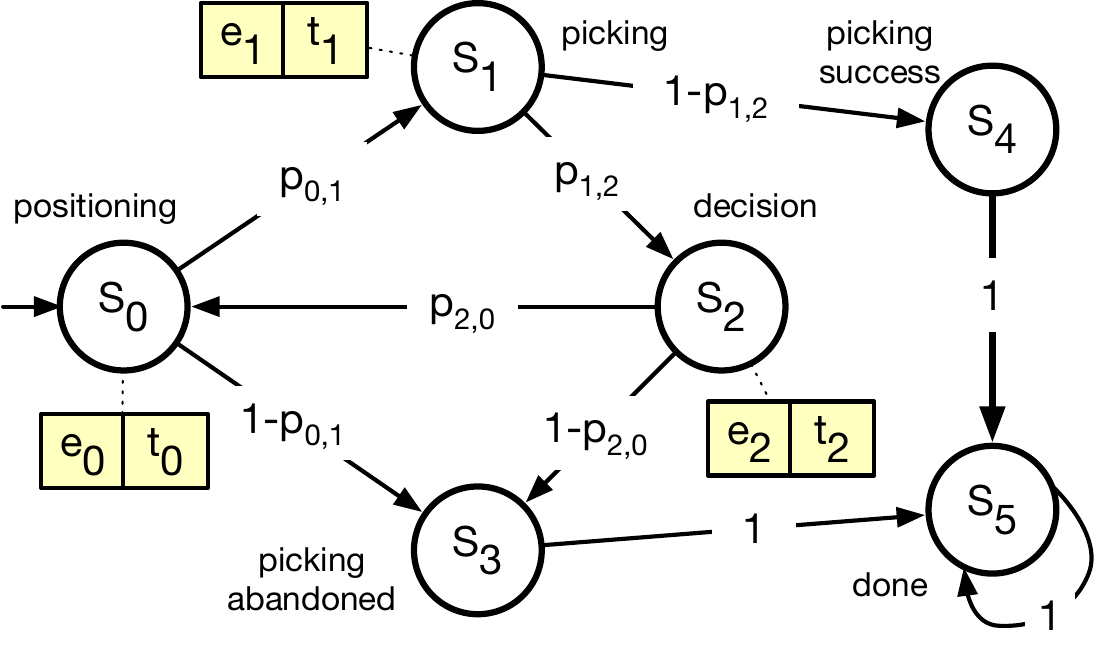}
	\vspace*{-4mm}
	\Description{DTMC of a robot executing a fruit-picking task; ($p_{i,j}$: unknown transition probabilities  between states.)}
	\caption{DTMC of a robot executing a fruit-picking task; ($p_{i,j}$: unknown transition probabilities  between states.)}
	\label{fig:dtmc}
	\vspace*{-2mm}
\end{figure}

Fig.~\ref{fig:dtmc} shows the DTMC model of this robotic fruit-picking task (FPR). 
Starting from state $s_0$, the robot takes the appropriate position (e.g., using its onboard camera and Lidar perception devices) close to the next fruit. 
The positioning operation succeeds with probability $p_{0,1}$, and the robot moves to state $s_1$, where it performs the picking operation. Conversely, the positioning operation fails with probability $1-p_{0,1}$, the robot moves to state $s_3$, and the task execution ends (state $s_5$). 
From state $s_1$, the picking operation is successful with probability $1\!-\!p_{1,2}$, the robot moves to state $s_4$ and then terminates its execution by moving to state $s_5$. 
In contrast, the picking is unsuccessful with probability $p_{1,2}$, and the robot moves to state $s_2$. 
In this state, the robot needs to decide whether to reposition itself (state $s_0$) and retry the entire process or to abandon the picking operation (state $s_3$) and end the process (state $s_5$);
this decision is shown probabilistically with transitions $p_{2,0}$ and $1-p_{2,0}$, respectively.

The DTMC model is annotated with two reward functions, illustrated as rectangular boxes linked to states $s_0$, $s_1$ and $s_2$ (Fig.~\ref{fig:dtmc}).
The ``time'' reward function associates mean operation execution times $t_0$, $t_1$ and $t_2$ with the three operations performed by
the robot.
Likewise, an ``energy'' reward function associates mean energy consumption $e_0$, $e_1$ and $e_2$ with the same operations. 
Finally, %we consider that 
the robot must \changed{validate} the three system-level requirements from Table~\ref{tab:prestoReqs}.

\begin{table}[t]
	\centering
	{\small
		\renewcommand{\arraystretch}{1.2}
		\caption{System-level requirements and formalised temporal logic properties for the fruit-picking robot.% 
		}\label{tab:prestoReqs}
		\vspace*{-4mm}
		\begin{tabular}{p{0.05cm}p{5.3cm}p{2cm}}%p{0.5cm}}
		\hline
		\textbf{$\!\!\!$ID} 
		& $\!\!\!$\textbf{Description}
		& $\!\!\!$\textbf{PCTL}\\
		% & $\!\!\!$\textbf{Type}\\
		\hline
		$\!\!\!$R1 
		&$\!\!$What is the probability for completing the fruit-picking process successfully? %with a probability of at least 80\%
		&$\!\!\!$${\sf P_{=?} [F\;\textrm{``success"}]}$ \\
		% &PK\\
		$\!\!\!$R2 
		&$\!\!$What is the expected time for completing the fruit-picking process?% shall not exceed 30 seconds.
		&$\!\!\!$${\sf R^{``time"}_{=?} \![F \textrm{``done"}]}$\\
		% &PK\\
		$\!\!\!$R3 
		&$\!\!$What is the expected energy consumption for completing the fruit-picking process?% shall not exceed 10 joules.
		&$\!\!\!$${\sf R^{``energy"}_{=?} \! [F \textrm{``done"}]}$
		% &$\;$E
		\\[1mm]
		\hline
	\end{tabular}
}
\vspace*{-4mm}
\end{table}

\section{\approach}\label{sec_method}
%\vspace*{-1mm}

\begin{figure*}[t]
    \centering\includegraphics[width=0.95\linewidth]{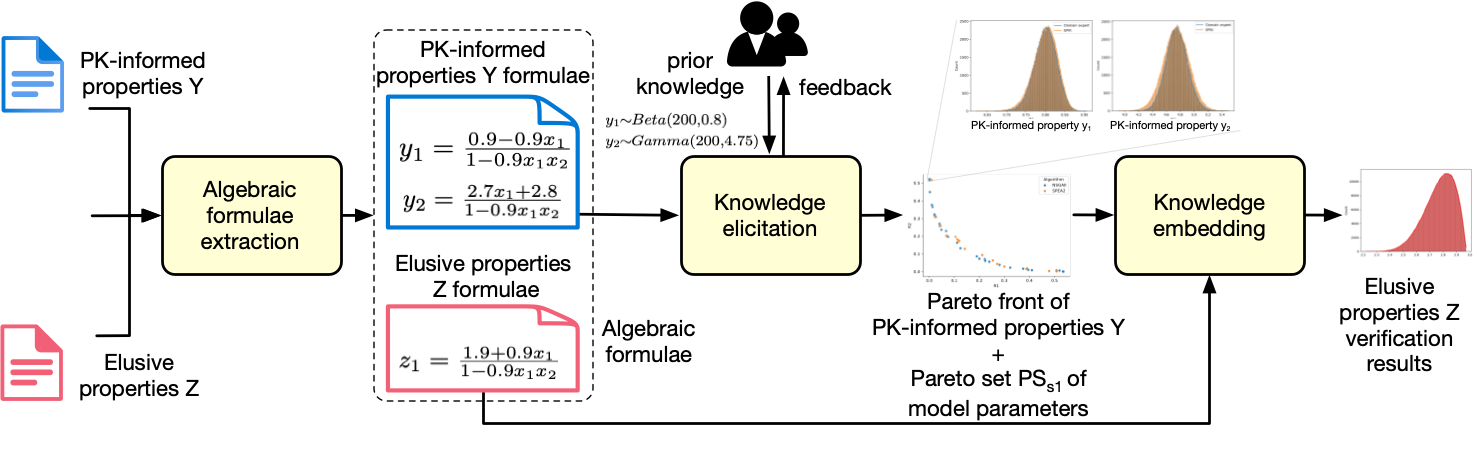}
    \vspace{-4mm}
    \Description{High-level EPIK workflow showing its main stages for algebraic formulae extraction of system-level properties via parametric model checking, and knowledge elicitation and embedding via solving a multi-objective optimisation problem.}
    \caption{High-level EPIK workflow showing its main stages for algebraic formulae extraction of system-level properties via parametric model checking, and knowledge elicitation and embedding via solving a multi-objective optimisation problem.}
    \label{fig:epik}
    \vspace{-2mm}
\end{figure*}

\subsection{Problem Formulation}
\label{sec_prob_formulation}

Although the canonical parameterisation of priors (e.g., the Beta in Eq.~\eqref{eq_prior_beta}) 
provides an intuitive mechanism for incorporating prior knowledge from domain experts and historical data, it still targets individual parameters of a Markov model. 
Such formal abstraction makes it challenging for experts to specify and accurately express the knowledge from historical data. 
For instance, specifying the probability $p_{1,2}$ of failing to complete the picking operation entails a detailed understanding and expertise of the Markov model underpinning the FPR system.
To address this issue, \approach\ incorporates a novel concept that simplifies the elicitation of prior knowledge.
Instead of requesting experts to express knowledge about model transition parameters, we ask them to express their knowledge about the \textit{observable system properties}. 
Hence, \approach\ seamlessly integrates with existing Bayesian estimators~\cite{epifani_change-point_2010,zhao_interval_2020,filieri_lightweight_2015} and strengthens the practical use of Bayesian learning for probabilistic model checking.

\begin{definition}[PK-informed and Elusive Properties]
We categorise system-level properties into two types: 
\\\noindent
$\bullet$ \textbf{PK-informed properties} for which domain experts possess (substantial) prior knowledge (PK) often gained via empirical observations of their value distributions in past executions of the same system or other systems exhibiting similar behaviour;
\\\noindent
$\bullet$ \textbf{Elusive properties} signifying properties that are unusual or novel or even properties that are expensive or risky to observe, making them challenging to gather knowledge about.
\end{definition}

Domain experts may feel more comfortable and be more willing 
to provide prior knowledge for properties of the former type. 
However, % despite the challenges involved, 
acquiring an understanding of elusive properties is crucial for establishing the overall trustworthiness of the subject system. 
% from a verification standpoint.

\begin{example}[PK-informed versus Elusive Properties]
%\vspace{-2mm}
\label{exp_two_types_properties}
Consider the fruit-picking robot (Section~\ref{sec_example}) and its formalised requirements (Table~\ref{tab:prestoReqs}).
Assume that domain experts have accumulated rich prior knowledge about requirements R1 and R2 from historical robotic missions, i.e., ``based on the past (previous and similar) 200 FPR missions, we estimate that the average probability of a successful mission is 0.8 and the expected mission time is 4.75 seconds''.
However, no battery usage information was recorded, signifying that no knowledge about requirement R3 is available. 
In this scenario, R1 and R2 are PK-informed properties, and R3 is the elusive property.
%\vspace*{-1mm}
\end{example}

We introduce the notations: $X\!\!=\!\!\{ x_1, ..., x_K\}$ is the set of unknown model transition parameters, $Y\!=\!\{ y_1, ..., y_I\}$ is the set of PK-informed properties and $Z\!=\!\{ z_1, ...,z_J\}$ is the set of elusive properties, with cardinality $K$, $I$ and $J$, respectively. 

For each PK-informed property $y_i\!\in\! Y$, we assume that domain experts define a probability distribution $PK_i(y_i)$ as their prior knowledge. 
Each unknown parameter $x_k$ can be estimated as a probability distribution $d_k(x_k;\theta_k)$ where $\theta_k$ is the vector of parameters characterising distribution $d_k$.
For instance, for a Beta distribution, $\theta_k$ is a vector comprising the two parameters from Eq.~\eqref{eq_prior_beta}, used in~\cite{epifani_model_2009} for transition probabilities, and, similarly, a Gamma distribution for transition rates~\cite{filieri_formal_2012}. 
The boldface $\pmb{\theta}\!=\![\theta_1,\!\dots\!,\theta_K]$ defines a vector. 

\begin{example}[PK-Informed Properties]
\label{exp_pk_as_distribution}
We build on Example~\ref{exp_two_types_properties} where R1 and R2 are PK-informed and R3 is the elusive property, denoted as $y_1$, $y_2$ and $z_1$ respectively (i.e., $Y=\{y_1,y_2\}$ and $Z=\{z_1\}$). 
Then, using the experts' knowledge ``based on previous and similar 200 fruit-picking missions, we know the average probability of a successful mission is 0.8 and the expected mission time is 4.75 seconds'', we can formalise the PK-informed properties $PK_{R1}$ and $PK_{R2}$, using the canonical parameterisation, as $y_1 \sim \mathit{Beta}(200,0.8)$ and $y_2 \sim \mathit{Gamma}(200,4.75)$.
Since $y_1$ is a probability and $y_2$ is a reward, we use $\mathit{Beta}$ and $\mathit{Gamma}$, respectively, to match their support $[0,1]$ and $[0,+\infty]$. 
% %\vspace*{-1mm}
\end{example}

\subsection{Algebraic Formulae Extraction}
\label{sec_formulae_extraction}
\approach, whose high-level workflow is shown in Fig.~\ref{fig:epik}, comprises three stages: 
first, extracting algebraic expressions for the PK-informed $Y$ and elusive $Z$ properties; 
then, eliciting knowledge about transition parameters $X$ through the PK-informed properties $Y$; 
and, finally, embedding the derived knowledge to verify the elusive properties $Z$.

The first \approach\ stage leverages ParaMC (Section~\ref{sec_preliminaries}) to derive closed-form rational functions $f_i$ and $g_j$ for the PK-informed $Y$ and elusive properties $Z$ such that $y_i=f_i(x_1,\!\dots\!,x_K), \forall i=1..I$, and $z_j=g_j(x_1,\!\dots\!,x_K), \forall j\!=1..J$.  
Extracting these algebraic formulae facilitates the accelerated execution of the other \approach\ stages. 
Since we can transform the knowledge elicitation and embedding stages into optimisation problems that can be solved independently, we avoid the repeated invocation of the PMC in the loop during the optimisation process. 
\changed{By avoiding the expensive PMC invocation, \approach\  avoids both the concrete model construction and model verification against the selected set of $Y$ PK-informed properties for each parameter instantiation.}
If, however, extracting the algebraic formulae is impossible due to the complexity of the Markov model or the property~\cite{fang2023fast,jansen_accelerating_2014}, invoking the probabilistic model checker (e.g., PRISM) iteratively would yield the same outputs, albeit much slower.
\changed{Accordingly, \approach's algebraic formulae extraction step is recommended, as it enables faster execution of knowledge elicitation described next, but it is not mandatory. 
Nevertheless, if a probabilistic model checker is used in the loop, \approach\ would operate equally well and yield the same output.}

\begin{example}[FPR Formulae Extraction]
\label{exp_pmc_result}
Assume $p_{1,2}$ and $p_{2,0}$ of the FPR DTMC (Fig.~\ref{fig:dtmc}) are the unknown transition parameters denoted as $x_1$ and $x_2$ (i.e. $X\!=\!\{x_1,x_2\}$), with known parameters $p_{0,1}\!=\!0.9$, $e_0\!\!=\!\!e_1\!\!=\!\!e_2\!=\!1$, $t_0\!\!=\!\!1$, $t_1\!\!=\!\!2$ and $t_2\!\!=\!\!3$. 
% Then, the algebraic expressions derived from employing parametric model checking~(Section~\ref{paraMC}) are 
Then, the derived algebraic formulae are:%from ParaMC~(Section~\ref{paraMC}) are 
% $$y_1=f_{R1}(x_1,x_2)=\frac{0.9-0.9x_1}{1-0.9x_1 x_2}$$
% $$y_2=f_{R2}(x_1,x_2)=\frac{2.7x_1+2.8}{1-0.9x_1 x_2}$$ 
% $$z_1=g_{R3}(x_1,x_2)= \frac{1.9+0.9x_1}{1-0.9x_1 x_2}$$
\vspace{-2mm}
$$y_1=f_{R1}(x_1,x_2)=(0.9-0.9x_1)/(1-0.9x_1 x_2)$$
\vspace{-6mm}
$$y_2=f_{R2}(x_1,x_2)=(2.7x_1+2.8)(1-0.9x_1 x_2)$$ 
\vspace{-5mm}
$$z_1=g_{R3}(x_1,x_2)= (1.9+0.9x_1)(1-0.9x_1 x_2)$$
\noindent
The unknown transition parameters $x_1$ and $x_2$ conform to Beta distributions as commonly used~\cite{epifani_model_2009} with optimisable parameters $\pmb{\theta}=[\theta_1,\theta_2]$ 
such that 
% $x_1 \sim Beta( \theta_1)$, $\quad x_2 \sim Beta( \theta_2)%\vspace{-2mm}$  
%\vspace{-2mm}$$x_1 \sim Beta( \theta_1), \quad x_2 \sim Beta( \theta_2)%\vspace{-2mm}$$  
$\theta_1$ is a vector of the two Beta distribution parameters, i.e., $\theta_1=[ n_1^{(0)},p_{1,2}^{(0)}]$ and $\theta_2=[ n_2^{(0)},p_{2,0}^{(0)}]$ (cf. Eq.~\eqref{eq_prior_beta}).
\end{example}

% \begin{example}[Parametric Distributions of Unknown Parameters to be Optimised]
% \label{exp_para_dist_x}
%     For the two unknown transition parameters $x_1$ and $x_2$, we assume that they conform to Beta distributions as commonly used in~\cite{epifani_model_2009,epifani_change-point_2010,zhao_probabilistic_2019} with optimisable parameters $\pmb{\theta}=[\theta_1,\theta_2]$ such that
%     $$x_1 \sim Beta( \theta_1), \quad x_2 \sim Beta( \theta_2)$$ 
%     where $\theta_1$ is a vector of the two Beta distribution parameters $\theta_1=[ n_1^{(0)},p_{1,2}^{(0)}]$, and similarly $\theta_2=[ n_2^{(0)},p_{2,0}^{(0)}]$ (cf. Eq.~\eqref{eq_prior_beta}).

% \end{example}

\subsection{Knowledge Elicitation}
\label{sec_knowledge_elicitation}
% %\vspace{1mm}\noindent
% \textbf{Elicitation Stage.} 
During this stage, \approach\ aims to find the values $\theta_1,..,\theta_K$ that best approximate the distribution of the unknown model parameters $x_1,\dots,x_K$ such that the distance between the properties distribution provided by domain experts and the distribution produced using $\theta_1,..,\theta_K$ is minimised. Formally, 
\begin{equation}
\label{eq_obj}
{%\tiny
	\min_{\substack{\theta_1 \\\dots \\ \theta_K}} \!\! \left\{D_{K\!L}\!\Bigl(PK_i(Y_i) || \mathbb{P}_{ \!\!\!\substack{X_1 \sim d(x_1;\theta_1) \\ \dots \\ X\!_K \sim d(x_K;\theta_K)}} \!\!\! 
	f_i(X_1 ,.., X_K) \!\!=\!\!Y_i   \Bigr), \!\! \forall i=1..I 
	\right\}
}
\end{equation}
where $\!D_{KL}(P||Q)\!$ is the Kullback–Leibler (KL) divergence~\cite{KLdivergence} 
% \footnote{Other distribution distance metrics may also serve the task.  We choose the KL divergence for its asymmetric property, giving the distance of the estimated distribution against the ground truth.} 
% i.e., optimising one distribution to the other, the ground truth one.} 
of the probability distributions $P$ and $Q$, %which is 
often used in information theory and interpreted as the amount of information lost when using $Q$ to approximate $P$.
% Note, for simplicity, 
Here we present the case where the $K$ unknown transition parameters $x_1,\!..\!,x_K$ are independent, assuming they are outgoing transition parameters from $K$ different states of a Markov model. 
If, however, $x_1$ and $x_2$ are outgoing transition parameters from the same DTMC state, then they cannot be assumed to be independent. 
Instead a joint prior distribution like a $\mathit{Dirichlet(x_1,\!x_2,\!1\!\!-\!\!x_1\!\!-\!\!x_2;\alpha_1,\alpha_2,\alpha_3)}$ (where $\theta_1$ and $\theta_2$ collectively become $\alpha_1,\alpha_2,\alpha_3$) should be used. 
\approach\ naturally supports such cases of optimising joint prior distributions.

\changed{
While KL divergence aligns naturally with information-theoretic objectives, alternative statistical divergence measures such as Jensen-Shannon (JS) divergence or Wasserstein distance can be employed to further support or refine knowledge elicitation~\cite{cai2022distances}. 
For instance, using the JS divergence offers a symmetric, bounded alternative that eliminates numerical instabilities in regions where the support of the transition distributions does not perfectly overlap. 
Alternatively, the Wasserstein distance (Earth Mover's Distance) could be integrated to exploit the underlying geometric structure of the state space, providing stable, non-vanishing gradients even when comparing disjoint probability distributions. 
Performing a robustness analysis of these alternative measures with respect to the accurate extraction of the $K$ transition parameters $X$, particularly under varying constraints of data sparsity or noise, is an interesting future work direction.
}

%When $I=1$, Eq.~\eqref{eq_obj} becomes a single-objective optimisation problem (SOOP) with a single optimal solution. 
%In general, 
For multiple PK-informed properties ($I>1$),  Eq.~\eqref{eq_obj} produces a multi-objective optimisation problem (MOOP) that involves finding the \textit{Pareto-optimal set} of solutions $PS_{s1}\!\!=\!\!\{ \pmb{\theta} \in \Theta^K | \pmb{\theta}=[\theta_1,\dots,\theta_K]\}$ 
that are on the Pareto front of the MOOP.
% \footnote{
The Pareto front is a well-established concept~\cite{pareto1964cours} describing an optimal solution set where each solution cannot be improved further in one objective without worsening its performance in another objective.
% } 
%
\begin{example}[Elicitation Stage]
\label{exp_moop_soop}
Following from the previous examples, %(and reusing the notations),
%where R1 and R2 ate the two elements in $Y$ and assuming $p_{1,2}$ $p_{2,0}$ in Fig.~\ref{fig:dtmc} are the two unknown transition parameter $x$, 
the MOOP of \approach's elicitation stage is given by: 
%\vspace{-2mm}
\begin{align}
\label{eq_obj_example_MOOP_SOOP_stage1}
&\min_{n_1^{(0)},p_{1,2}^{(0)},n_2^{(0)},p_{2,0}^{(0)}}  \nonumber
\\[-2mm]
\Biggl\{
& D_{KL}\Bigl(PK_{R1}(Y_1) || \mathbb{P}_{\substack{x_1 \sim \mathit{Beta}(n_1^{(0)},p_{1,2}^{(0)})\\ x_2 \sim \mathit{Beta}(n_2^{(0)},p_{2,0}^{(0)})}}
(f_{R1}(x_1,x_2) =Y_1)   \Bigr),  \nonumber
\\[-4mm]
% %\vspace{-8mm}
% \[-2mm\]
& D_{KL}\Bigl(PK_{R2}(Y_2) || \mathbb{P}_{\substack{x_1 \sim \mathit{Beta}(n_1^{(0)},p_{1,2}^{(0)})\\ x_2 \sim \mathit{Beta}(n_2^{(0)},p_{2,0}^{(0)})}}
(f_{R2}(x_1,x_2) =Y_2)   \Bigr) 
\Biggr\}
\end{align}
%\vspace{-6mm}
where  $PK_{R1}(Y_1)$ and $PK_{R2}(Y_2)$ are the distributions of the PK-informed properties.
\end{example}
% %\vspace{-1mm}

\begin{algorithm}[t!]
\small
\caption{\approach\ Elicitation Stage}
\label{alg_eliciting_opt}
\SetKwInput{KwInput}{Input}                % Set the Input
\SetKwInput{KwOutput}{Output}              % set the Output
\DontPrintSemicolon

\SetKwFunction{Fyfx}{y\_dist\_from\_x}
\SetKwFunction{Fobjs}{objectives}

\KwInput{
% \\$M$: parametric Markov chain model of the system with $x_k, \forall\, k=1 \ldots K,$ unknown model parameters
% \\$R_i, \forall i=1 \ldots I$: system requirements formalised in temporal logic (PCTL/CSL)
\\ $PK_i(y_i)$,$\forall i\!=\!1\!\ldots\!I$: 
PK-informed properties prior knowledge
\\$f_i$, $\forall i\!=\!1\!\ldots\!I$: 
PK-informed properties algebraic expressions 
% the ParaMC results $f_i, \forall i=1..I$; 
\\$d_k(x_k;\!\theta_k)$,$\forall k\!=\!1\!\ldots\!K$: 
distributions for %the $K$ 
model parameters
\\$n$: sample size 
}
\KwOutput{
\\$PF_{s1}=\{ objectives(\pmb{\theta}) \in \mathbb{R}^I \; \forall \,\pmb{\theta} \in PS_{s1}\}$
\\$PS_{s1}=\{ \pmb{\theta} \in \Theta^K | \pmb{\theta}=[\theta_1,..,\theta_K] \text{ is Pareto-optimal} \}$ 
% \hfill\tcp{\footnotesize Pareto-optimal approximation set} 
}
% Set Function Names
\SetKwFunction{FMain}{ElicitKnowledge}
\;
\SetKwProg{Fn}{Function}{:}{\KwRet}
\Fn{\FMain{...}}{
$PF_{s1} \leftarrow \emptyset$,
$PS_{s1} \leftarrow \emptyset$ 
\;
% $(f_1, \ldots, f_I) \leftarrow \textsc{ParaMC}(M, (R_1, \ldots, R_I))$ 
% \hfill\tcp{\footnotesize extract algebraic expression for each $R_i$}
\While{$\neg\textsc{Terminate}(PS_{s1}, \pmb{\theta})$}{
	$\pmb{\Theta}' \leftarrow \textsc{GetCandidateParamVectors}(\pmb{\theta}, PS_{s1})$\;
	\ForEach{$\pmb{\theta}' \in \pmb{\Theta}'$}{
		$OBJS_{\theta'} \gets \emptyset$\;
		\For{$i=1$ \KwTo $I$} { 
			sample$_i \gets \emptyset$\;
			\For{$k=1$ \KwTo $K$} {   
				sample$_{x_k} \gets$ \textsc{Random$(d_k(x_k;\theta_k),n)$}\;
				sample$_i \gets$ sample$_i \frown$ sample$_{x_k}$
			}
			sample$_y \gets f_i(\mathrm{sample}_{i1}, \ldots, \mathrm{sample}_{ik})$\;
			$\mathrm{fitted}_{PK_i} \gets \textsc{FitDistribution}(\mathrm{sample}_y$)\;
			$obj \gets \textsc{GetKL}(PK_i(y_i), \mathrm{fitted}_{PK_i}) $\;
			$OBJS_{\theta'}  \gets OBJS_{\theta'}  \frown obj$
		}
		$\!\!\mathrm{dom}_{OBJS_{\theta'}} \!\!\gets\!\textsc{GetDominated}(OBJS_{\theta'},\! PF_{s1})$\;\!\!\!
		% \If{$\textsc{Dominates}(OBJS_{\theta'}, PF_{s1})$}{%
			% $PF_{s1} = PF_{s1} \setminus \textsc{Dominated}(OBJS_{\theta'}, PF_{s1})$\;
			% $PF_{s1} = PF_{s1} \cup \{OBJS_{\theta'}\}$\;
			% $Q^* \gets Q^* \wedge t_i$\;
			% }
		\If{$|\mathrm{dom}_{OBJS_{\theta'}}| > 0$}{
			$PF_{s1} \!=\! PF_{s1} \setminus \mathrm{dom}_{OBJS_{\theta'}}$\;
			$PS_{s1} \!=\! PS_{s1} \setminus \{\pmb{\theta} \;|\;  OBJS_{\theta} \in \mathrm{dom}_{OBJS_{\theta'}}\}$\;      
		}
		\If{$|\textsc{GetDominated}(PF_{s1}, OBJS_{\theta'})| = 0$}{
			$PF_{s1} = PF_{s1} \cup \{OBJS_{\theta'}\}$\;      
			$PS_{s1} = PS_{s1} \cup \{\pmb{\theta}'\}$\;      
		}   
	}
	$PF_{s1}, PS_{s1} = \textsc{Diversify}(PF_{s1}, PS_{1})$
}
\KwRet $PF_{s1}, PS_{s1}$\;
}

\end{algorithm}
% %%\vspace{-4mm}

We solve the MOOP from Eq.~\eqref{eq_obj} by transforming it into a search-based optimisation problem and leveraging the capabilities of evolutionary algorithms. 
At the core of the solution lies a multi-objective genetic algorithm driving the synthesis of the Pareto Front $PF_{s1}$ and its corresponding Pareto set $PS_{s1}$.
Algorithm~\ref{alg_eliciting_opt} shows the high-level steps of the solution which receives as inputs the distributions of the PK-informed system properties based on the prior knowledge from domain experts, the algebraic expressions for these properties and information related to the distribution type of each unknown model transition parameter $\theta_k$.
Then, the evolutionary loops are executed (lines 4--25) that involve the generation of candidate solutions (line 5) and the evaluation of each solution (lines 6--23) during which the solution values are used to instantiate the distribution of unknown model parameters (lines 10--13); distribution fitting occurs for each $f_i$ outcome (line 14) and the KL divergence is calculated for all PK-informed properties (lines 15--16).
Next, the normal Pareto dominance is executed to remove inferior solutions and add non-dominated solutions (lines 17--23).
The \textsc{Diversify} function (line 24) employs evolutionary algorithm strategies for diversity preservation to reduce the possibility of premature convergence and also to select the solutions that will participate in the next generation. 
Once the loop terminates, the Pareto-optimal solutions set $PS_{s1}$ 
and its corresponding Pareto front $PF_{s1}$ are returned.

% In Algorithm~\ref{alg_eliciting_opt}, we first declare global variables for storing all the required inputs. Then a function is declared at line 4-9, which takes the distribution parameters $\pmb{\theta}$ of all the $K$ unknown transition parameters $X$ and returns the fitted distribution of a single PK-informed property $y_i\in Y$. From line 11-16, we try to define a set of objective functions for the later MOOP. Each objective, regarding a single PK-informed property, represents the KL divergence between the prior knowledge given from experts and the fitted distribution from (unknown) model transition parameters after PMC. Collectively, these $I$ objectives (corresponding to $I$ PK-informed properties) forms the set of objectives in the MOOP at line 21 where $\pmb{\theta}$ is the set of variables to be optimised. After invoking some numerical optimisation tool at line 21, the Pareto-optimal set of solutions $PS_{s1}$ and its corresponding optimised objectives $objs^*$ are returned.

\subsection{Knowledge Embedding}
\label{sec_knowledge_embedding}
% %%\vspace{1mm}\noindent
% \textbf{Embedding Stage.} 
Given the Pareto-optimal set of solutions $PS_{s1}$ as the outcome of the elicitation stage, \approach\ executes the knowledge embedding stage to establish the verification results for the elusive properties $Z$ and select the most appropriate solution from $PS_{s1}$. 
To achieve this, \approach\ solves another optimisation problem. 
If $J\!=\!1$ (one elusive property), the problem is single-objective; otherwise (if $J\!>\!1$), the problem is multi-objective, meaning that a reduced Pareto set $PS_{s2} \subseteq PS_{s1}$ will be derived. 
Deciding the estimates (e.g., quantiles, moments) to use and solve the optimisation problem should align with the expectations of decision-makers. 
Similarly, the adoption of a conservative or optimistic stance prescribes whether optimising the elusive properties $Z$ involves their maximisation or minimisation. 
% A similar reasoning holds for the adoption of  a conservative or optimistic stance, which prescribes whether optimising the elusive properties $Z$ involves their maximisation or minimisation. 
Without loss of generality, the following formulation assumes the minimisation of the expected value for all elusive properties: 
\begin{equation}
\label{eq_embed_moop_stage_2}
\min_{[\theta_1,..,\theta_K] \in PS_{s1}} \left\{ \mathbb{E}_{ \substack{X_1 \sim d(x_1;\theta_1) \\ \dots \\ X_K \sim d(x_K;\theta_K)}}[g_j(X_1,\dots, X_K)] , \forall j=1..J \right\}
\end{equation}
% which is a SOOP when $J=1$ (i.e., there is only one elusive property to verify), and an MOOP otherwise that involves finding the Pareto optimal set $PS_{s2} \subseteq PS_{s1}$.  
%
The algorithm for extracting the Pareto set $PS_{s2}$ and its corresponding Pareto front $PF_{s2}$ is similar to Algorithm~\ref{alg_eliciting_opt}.
The key difference pertains to the main loop (lines 4--5) where the \textsc{GetCandidateParamVectors} function retrieves solutions from $PS_{s1}$ (instead of synthesising new) and the \textsc{Terminate} function holds when all solutions in $PS_{s1}$ have been examined. 
We omit this algorithm for brevity reasons.

\begin{example}[Embedding Stage]
\label{exp_moop_soop_2}
Given the Pareto-optimal set from the elicitation stage (Example~\ref{exp_moop_soop}) and elusive property R3 that will be maximised (for conservative purposes), the single-objective problem during \approach's embedding stage is given by:
\begin{equation}
%\vspace{-3mm}
\label{eq_obj_example_MOOP_SOOP_stage2}
\max_{[n_1^{(0)},p_{1,2}^{(0)},n_2^{(0)},p_{2,0}^{(0)}] \in PS_{s1}} \mathbb{E}_{\substack{x_1 \sim \mathit{Beta}(n_1^{(0)},p_{1,2}^{(0)})\\ x_2 \sim \mathit{Beta}(n_2^{(0)},p_{2,0}^{(0)})}}
[g_{R3}(x_1,x_2)] 
\end{equation}
%\vspace{-2mm}
\end{example}

\vspace{1mm}\noindent 
%\simos{This is true only when using algebraic expressions for the \approach. Otherwise, the verification time per property depends on the model and the model checker.}
\textbf{Complexity Analysis.}  
In \approach's elicitation stage (Algorithm~\ref{alg_eliciting_opt}), since the time complexity of simple random sampling is generally $O(n)$, lines 10--12 together yield time complexity $O(n*K)$.
Since $f_i$ is a closed-form rational function with $O(1)$ and each sample has a size of $n$, the time complexity of line~13 is $O(n)$. 
Regarding \textsc{FitDistribution} (line~14), while the time complexity for distribution fitting can vary (it depends on the specific algorithm and the characteristics of the distribution being fitted), it is often\footnote{Specifically, in Section~\ref{sec_eva}, we first calculate the mean (in $O(n)$ time complexity) and variance (in $O(n)$ time by calculating the sum of squared differences, in addition to the mean) of the sample with size $n$, then derive the two distribution parameters for Beta/Gamma from them (in $O(1)$ constant time).
Thus, the total time complexity is $O(n) + O(n) +O(1) = O(n)$.}
represented as $O(n)$ where $n$ is the number of data points used for the fitting. 
The time complexity of the KL divergence calculation (line~15) $D_{KL}(P||Q) = \sum_{i}P(i)\log\frac{P(i)}{Q(i)}$ is $O(n)$ by definition, due to the need to iterate through all $n$ elements to compute $P(i)$ and $Q(i)$.
Since this is executed for all $I$ PK-informed properties, the total time complexity is for evaluating a candidate solution \pmb{$\theta'$} is $O(n*K*I)$. 
Concerning the evolution loop itself, this is computationally intensive and the complexity really depends on the chosen evolutionary algorithm. 
For example, \approach\ instrumented with a multi-objective genetic algorithm has a general complexity $O(I*G*S^2)$ where $I$ is the number of objective functions, $S$ is the population size and $G$ is the number of generations\footnote{A genetic algorithm like NSGA-II~\cite{deb2002fast} has time complexity $O(I*S^2)$ per generation~\cite[Chapter 10.4]{Sastry2005}. Over $G$ generations, the total time complexity is $O(G*I*S^2)$.}. 
Consequently, the time complexity of the whole Algorithm~\ref{alg_eliciting_opt} to the size of various inputs is $O(I^2*K*n*G*S^2)$.

The key differences to note for \approach's knowledge embedding algorithm are: 
(1) the set of objectives becomes a set of statistics (that are of practical interest in the given application) on the distributions of the $J$ elusive properties as encoded in Eq.~\eqref{eq_embed_moop_stage_2}; and
% Without loss of generality, at line 13, we illustrate Mean as an example; 
(2) the search space of the multi-objective optimisation problem is the result of the elicitation stage, i.e., the variables to be optimised are constrained in $PS_{s1}$.
Accordingly, the time complexity is $O(J^2*K*n*G*S^2)$; since similar time analysis steps apply for this algorithm too, we omit the details for brevity. 

\section{Evaluation}
\label{sec_eva}

\subsection{Research Questions}
\label{sec_rqs}
%\vspace{-2mm}
% We performed a comprehensive experimental evaluation to answer the following research questions:

\vspace{1mm}\noindent
\textbf{RQ1 (Accuracy): How closely can \approach\ approximate the true distribution of PK-informed properties?}
We analyse if \approach\ can derive the distributions of unknown transition parameters, thus yielding distributions of PK-informed properties that closely approximate the ground truth of those properties.

\vspace{1mm}\noindent
\textbf{RQ2 (Effectiveness): How do different optimisation approaches affect the effectiveness of \approach?}
% We used this research question to analyse 
We examine the impact of different multi-objective optimisation algorithms in \approach's performance. 
We study the quality of \approach-derived %unknown 
model transition parameters using the established evolutionary algorithms NSGA-II~\cite{deb2002fast},  SPEA2~\cite{zitzler2001spea2} and CMA-ES~\cite{igel2007covariance}.
% \\- Use 2-3 different optimisation algorithms (e.g., GA) and analyse their behaviour in terms of KL divergence and completion time
% \\- (if time/space is available) check their hyperparameters too.

\vspace{1mm}\noindent
\textbf{RQ3 (Conformance): How does using knowledge of PK-informed properties of varying levels of conformance affect \approach?}
% How effective is \approach\ with employing knowledge from PK-informed properties of varying levels of conformance?}
Deciding unknown transition parameters depends on the prior knowledge encapsulated in the set of PK-informed properties.
Since the encoded prior knowledge about these properties can be conflicting, with this research question, we examine its impact on \approach's convergence.

\vspace{1mm}\noindent
\textbf{RQ4 (Knowledge Embedding):  
Can \approach\ support verifying elusive properties for which prior knowledge is absent?}
Supporting decision-making entails providing useful insights into the trade-offs between transition parameter distributions elicited using \approach.
% produced during \approach\ knowledge elicitation. 
% To support decision-making and help software engineers make informed decisions, \approach\ must yield complementary solutions with different trade-offs. 
Thus, we explore how \approach\ supports the verification of elusive properties. %at design time 
% and their runtime analysis~\cite{epifani_model_2009,zhao2023bayesian}.
% evaluated these trade-offs for the systems used in our evaluation, also considering both how \approach\ can be used to verify elusive properties at design and support verification at runtime~\cite{epifani_model_2009,zhao2023bayesian,gerasimou_efficient_2014}.
% How can EPIK support the verification of new properties for which prior knowledge does not exist?}
% \\- Use the results from above to verify new properties for the use cases
% \\- Scenario 1: Assume that we have new knowledge for a particular property from someone expert joining the team. This new knowledge should be incorporated into the PK-informed property
% \\- Scenario 2: Assume that we have new knowledge for a new property from someone expert joining the team. This new property (with its knowledge) should be become a new PK-informed property and used to re-estimate the model parameters
% \\- Measure the KL divergence in these scenarios

\subsection{Evaluation Methodology}
\label{sec_eva_method}

\textbf{Software Systems.}
We evaluated \approach\ using several variants of two software-controlled systems from different application domains:
(1) the fruit-picking robot (FPR)~\cite{fang2022presto} (Section~\ref{sec_example}); and
(2) a service-based system for foreign exchange trading (FX) presented in~\cite{gerasimou_search-based_2015}.

\vspace{1mm}\noindent
\emph{FX Description (FX).}
An FX trader can use the system in two modes. 
Using the \emph{expert} mode, FX iteratively analyses market activity, identifies patterns that satisfy the trader's objectives, and automatically carries out trades.
To this end, a \emph{Market watch} operation extracts real-time exchange rates (bid/ask price) of selected currency pairs, which is used by a \emph{Technical analysis} operation to assess the current trading conditions, predict future price movement, and decide if the trader's objectives are: (i)~``met'' (causing the invocation of an \emph{Order} service to carry out a trade); (ii)~``not met'' (resulting in a new \emph{Market watch} invocation); or (iii)~an error occurred (triggering an \emph{Alarm} operation to notify the trader about discrepancies/opportunities not covered by the trading objectives).
Using the \emph{normal} mode, FX assesses the economic prospects of a country via a \emph{Fundamental analysis} operation that collects, analyses and evaluates information (e.g., news reports, economic data and political events), and provides an assessment on the country's outlook. 
If satisfied with this assessment, the trader can use the \emph{Order} operation to sell/buy currency; then, a \emph{Notification} operation confirms the trade completion.
% of the trade. 
Finally, FX engineers are interested in analysing the system-level requirements in Table~\ref{tab:fxReqs}.

\begin{table}[t]
\centering
{\small
\renewcommand{\arraystretch}{1.2}
\caption{FX system-level requirements and properties
%and formalised temporal logic properties for the FX system. 
% [`Type': PK-informed (PK)  or elusive (E) requirement]
}\label{tab:fxReqs}
\vspace*{-4mm}
\begin{tabular}{p{0.1cm}p{5.3cm}p{2cm}}%p{0.5cm}}
\hline
\textbf{$\!\!\!$ID} 
& $\!\!\!$\textbf{Description}
& $\!\!\!$\textbf{PCTL}\\
% & $\!\!\!$\textbf{Type}\\
\hline
$\!\!\!$R1 
&$\!\!$What is the probability that FX will complete the workflow successfully? %with a probability of at least 80\%
&$\!\!\!$${\sf P_{=?} [F\;\textrm{``success"}]}$ \\
% &PK\\
$\!\!\!$R2 
&$\!\!$What is the expected response time for completing workflow execution?% shall not exceed 30 seconds.
&$\!\!\!$${\sf R^{``time"}_{=?} \![F \textrm{``done"}]}$\\
% &PK\\
$\!\!\!$R3 
&$\!\!$What is the total cost of the third-party services used
for  a workflow execution?% shall not exceed 10 joules.
&$\!\!\!$${\sf R^{``cost"}_{=?} \! [F \textrm{``done"}]}$
% &$\;$E
\\[1mm]
\hline
\end{tabular}
}
\vspace{-3mm}
\end{table}

\begin{table}[t]
\centering
{\small
\renewcommand{\arraystretch}{1.2}
\caption{FPR and FX system variants analysed using \approach}\label{tab:systemVariants}
\vspace*{-4mm}
\begin{tabular}{p{0.6cm}p{6.2cm}p{0.9cm}}
% p{4.6cm}p{2.7cm}}
\hline
\textbf{$\!\!\!$Variant} 
& $\;$\textbf{Details}
& $\!\!$\textbf{Size}\\
\hline
%p1, p2, p3, R1, R2, R3
\changed{$\!\!\!$FPRc} 
& \changed{Y=\{R1\}; Z=\{R2,R3\}; X=\{$p_{2,0}$\}}
& \changed{1E+08}\\
$\!\!\!$FPR1 
& Y=\{R1,R2\}; Z=\{R3\}; X=\{$p_{2,0}$\}
& 1E+08\\
% &$R1(a_1, b_1), R2(a_2,b_2)$\\
$\!\!\!$FPR2 
& Y=\{R1,R2\}; Z=\{R3\}; X=\{$p_{1,2}, p_{2,0}$\}
& 1E+16\\
$\!\!\!$FPR3 
& Y=\{R1,R2\}; Z=\{R3\}; X=\{$p_{0,1}, p_{1,2}, p_{2,0}$\}
& 1E+24\\
$\!\!\!$FX1 
& Y=\{R1,R2\}; Z=\{R3\}; X=\{$p_{Or1}$\}
& 1E+8\\       
$\!\!\!$FX2 
& Y=\{R1,R2\}; Z=\{R3\}; X=\{$p_{Or1}, p_{F\!A2}$\}
& 1E+16\\
$\!\!\!$FX3
% &
& Y=\{R1,R2\}; Z=\{R3\}; X=\{$p_{Or1}, p_{F\!A2}, , p_{T\!A1}$\}
& 1E+24\\
% $\!\!\!$DPM\_1
%     &
%     &\\
% $\!\!\!$DPM\_2
%     &
%     &\\
% $\!\!\!$DPM\_3
%     &
%     &\\
[1mm]
\hline
\end{tabular}
}
\vspace{-5mm}
\end{table}

\vspace{2mm}\noindent
\textbf{Experimental Setup.}
We executed a diverse set of experiments using the FPR and FX  system variants from Table~\ref{tab:systemVariants}. 
The `Details' column indicates the PK-informed properties set ($Y$), the elusive properties set ($Z$) and the unknown model transition parameters ($X$).
The 'Size' column reports the analysis cost an exhaustive search would incur, assuming two-decimal precision for each double-valued parameter needed to encode an unknown parameter $x_k$.
We equipped the optimisation algorithm of \approach\ using the established multi-objective genetic algorithms NSGA-II~\cite{deb2002fast} and SPEA2~\cite{zitzler2001spea2} and CMA-ES~\cite{igel2007covariance}, \changed{a state-of-the-art derivative-free evolution strategy based on covariance matrix adaptation.
Unlike NSGA-II and SPEA2, that employ crossover and mutations, CMA-ES models the search space as a multivariate normal distribution and continuously learns the correlation between decision variables (parameters) and adapts its search to steer the population toward optimal regions.
All algorithms leverage the DEAP Python package~\cite{de2012deap}.
}
We also used the following configuration to evaluate our approach:
2,000 evaluations with an initial population of 100 individuals (i.e., 20 generations in total), and default values for single-point crossover probability $p_c = 0.9$ and uniform polynomial mutation probability $p_m =0.8$. 
We selected these values after carrying out a set of preliminary experiments and following the standard practice in the field of search-based software engineering~\cite{gerasimou_search-based_2015,harman2012search,arcuri2011practical}.
\changed{For problem variants with $\geq\!2$ objectives (PK-informed properties), i.e., $|Y|\!>\!2$, we use standard Pareto dominance \eqref{eq_obj} to construct the Pareto front approximation~\cite{coello2007evolutionary}.}

% Thus, we executed 30 independent runs per system variant from Table~\ref{tab:systemVariants}. and each
% multiobjective optimisation algorithm~\cite{arcuri2011practical}. 
% All the experiments were run on an Apple MacBook Pro with an M2 Max processor and 64GB of memory.

% \changed{For problem variants with multiple objectives (PK-informed properties), i.e., $|Y|>2$, \approach\ uses standard Pareto dominance \eqref{eq_obj} to construct the Pareto front approximation~\cite{coello2007evolutionary}.}
To reduce the potential impact of non-deterministic behaviour affecting \approach's performance  (e.g., when choosing the crossover point for the genetic algorithms), we adopted the established procedure in search-based software engineering~\cite{harman2012search}.
Thus, we performed 30 independent runs per system variant from Table~\ref{tab:systemVariants} and each multi-objective optimisation algorithm~\cite{arcuri2011practical}.
All experiments were run on a MacBook Pro with Apple M2 Max processor and 64GB of memory.

\vspace{2mm}\noindent
\textbf{Statistical Analysis.} 
Since the actual Pareto front for the real-world systems from our experimental evaluation is unknown and its exact computation is typically impossible, we used the standard practice~\cite{zitzler2008quality}.
For each system variant, we extract the reference front comprising the nondominated Pareto front approximation from all the runs across all \approach\ instances.
We used this reference front and the Pareto front
quality indicators below to quantify the ‘goodness of fit’ of Pareto front approximations produced by \approach. 
We use a boxplot to show the central tendency and distribution per indicator.

$\bullet$ \textit{$I_{HV}$ (Hypervolume)}: measures the objective space volume consumed by a Pareto front approximation compared to a reference front. 
It shows diversity and convergence, and is strictly Pareto compliant~\cite{zitzler2008quality}. 
Larger $I_{HV}$ values give better Pareto front approximations.

$\bullet$ \emph{$I_{\epsilon}$ (Unary additive epsilon)}: measures the minimum value needed by a Pareto front approximation to dominate the reference front.
$I_{\epsilon}$ shows convergence to the reference front and is Pareto compliant. 
Smaller $I_{\epsilon}$ values denote better Pareto front approximations.

$\bullet$ \emph{$I_{IGD}$ (Inverted Generational Distance)}: the Euclidean distance in the objective space between the Pareto front approximation and the reference front.
\emph{$I_{IGD}$} shows diversity and convergence to the reference front.
A smaller $I_{IGD}$ gives better Pareto %front 
approximations.
% signifies an “error measure”

\begin{figure*}[t]
    \centering\includegraphics[width=\linewidth]{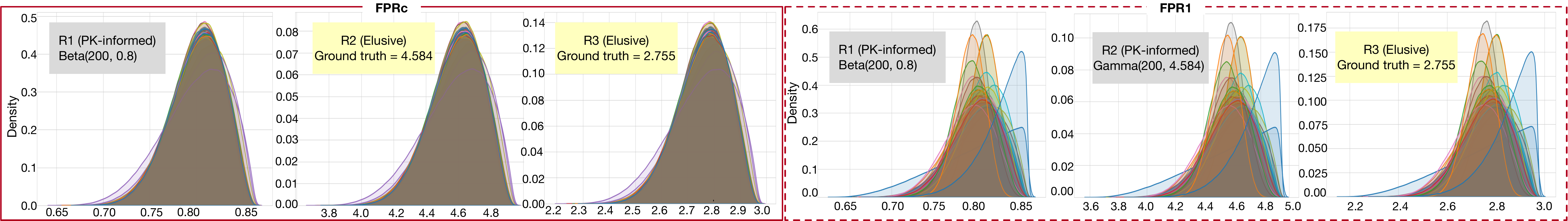}
    \vspace{-7mm}
    \caption{
    \changed{Distributions for the PK-informed and elusive properties for the FPRc controlled variant (1 PK-informed property) and FPR1 (2 PK--informed properties) over 30 independent runs, where ground truth R1=0.80, R2=4.584 and R3=2.755.}}
    \Description{
    \changed{Distributions for the PK-informed and elusive properties for the FPRc controlled variant (1 PK-informed property) and FPR1 (2 PK--informed properties) over 30 independent runs, where ground truth R1=0.80, R2=4.584 and R3=2.755.}}
    \label{fig_fpr_controlled}
    \vspace{-4mm}
\end{figure*}

Adopting the recommended practice~\cite{arcuri2011practical}, we employed inferential statistics to compare the quality indicator values obtained by \approach\ variants.  
Thus, we confirmed that the quality indicator values do not follow a normal distribution using the Shapiro-Wilk test.
Then, we used the Mann-Whitney and Kruskal-Wallis non-parametric tests with 95\% confidence level ($\alpha = 0.05$) to analyse the results without making assumptions about the data distribution or the variance homogeneity. 
Where appropriate, we did a post-hoc analysis with pairwise comparisons between the algorithms, using the conservative Bonferroni correction $p_{crit} = \alpha/k$ (k is the number of comparisons) to control the family-wise error rate.
When statistical significance exists, we use Cohen’s d to quantify the importance of the observed effect~\cite{arcuri2011practical}. 
Cohen’s d score summarises the difference between two groups as the number of standard deviations: $d=0.2$, $d=0.5$ and $d=0.8$ denote a small, medium and large effect size, respectively.

\subsection{Results and Discussion}
\label{sec_eva_results_disc}

\noindent
\textbf{RQ1 (Accuracy).}
\changed{First, we performed a controlled experiment to establish if the resulting estimates for elusive properties are accurate. 
Figure~\ref{fig_fpr_controlled} shows the resulting distributions for the PK-informed (grey box) and elusive (yellow box) properties for the FPRc (left) and FPR1 (right) variants over 30 independent runs.
The ground truth values for the elusive properties (R2=4.584 and R3=2.755 for FPRc; R3=2.755 for FPR1) were calculated using the PK-informed properties information, solving the system of linear equations to extract the expected value for the unknown model transition parameter $p_{2,0}$ and using the value for the elusive properties.
Then, we used \approach\ to extract values for the Beta distribution of $p_{2,0}$ and sampled from the distribution to construct the distribution of the PK-informed and elusive properties.
As shown, the derived distributions closely approximate the ground truth; the delta between the distributions' mean and (assumed) ground truth is $<0.00389$, signifying \approach's ability to retrieve meaningful and accurate elusive property estimates.}

\begin{figure}[t]
    \centering\includegraphics[width=\linewidth]{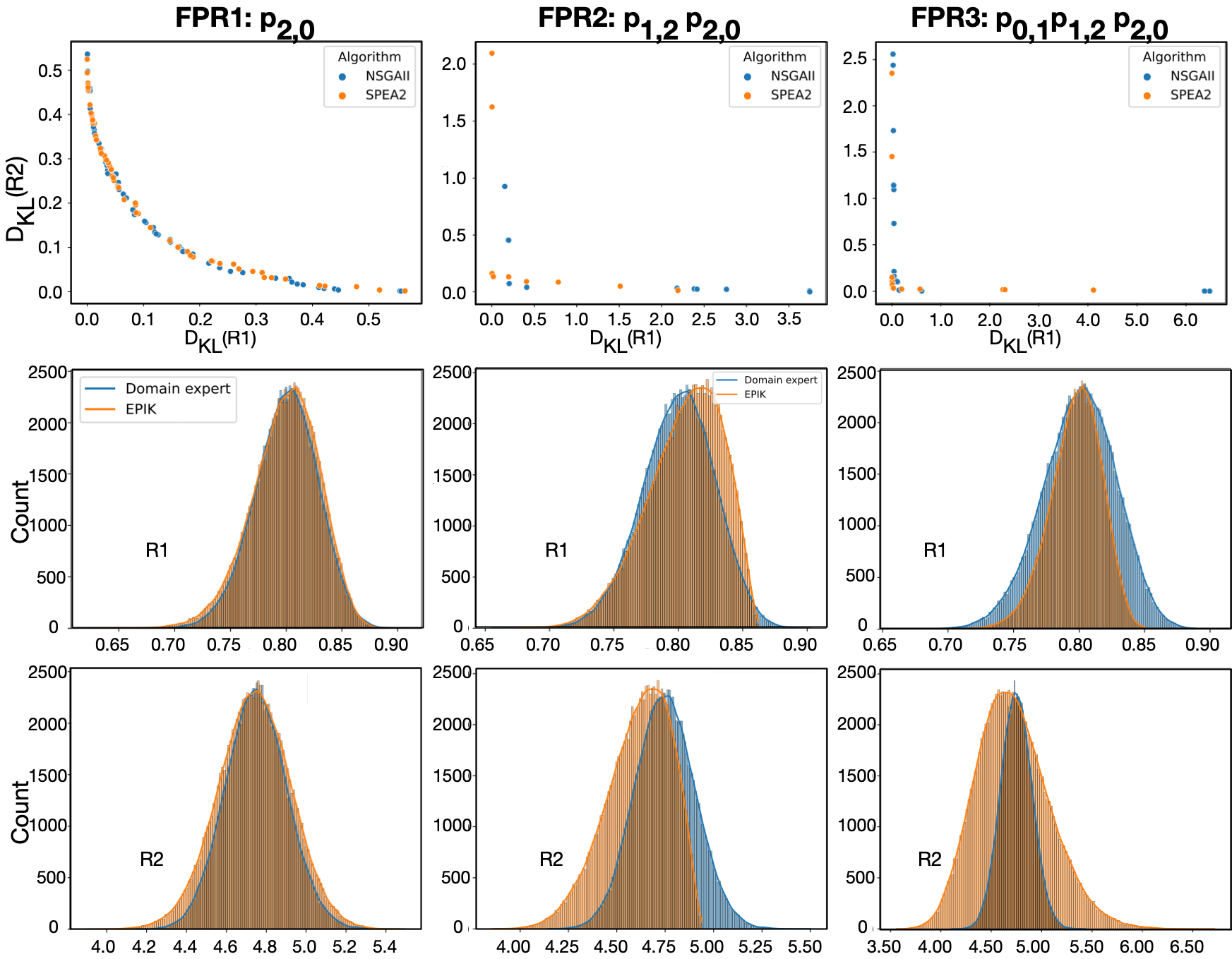}
    \vspace{-7mm}
    \Description{Pareto fronts (top) and sample distributions of requirements R1 (middle) and R2 (bottom) estimated by \approach\ for different numbers of unknown model transition parameters of the fruit-picking robot use case (variants FPR1, FPR2, FPR3)}
    \caption{Pareto fronts (top) and sample distributions of requirements R1 (middle) and R2 (bottom) estimated by \approach\ for different numbers of unknown model transition parameters of the fruit-picking robot use case (variants FPR1, FPR2, FPR3)}
    \label{fig_fpr_pf}
    \vspace{-6mm}
\end{figure}

Also, we establish whether \approach\ can accurately approximate the distributions of PK-informed properties by determining realistic distributions for the unknown transition parameters of the subject system models. 
Fig.~\ref{fig_fpr_pf} (top) shows example Pareto fronts for the FPR variants FPR1, FPR2 and FPR3 generated during the elicitation stage of \approach. 
Irrespective of the evolutionary algorithm underpinning our approach (NSGA-II or SPEA2), \approach\ is capable of producing Pareto front approximations with diversified KL divergence values ($D_{KL}$) for the considered PK-informed properties R1 and R2. 
In particular, the Pareto fronts for FPR1 comprise many non-dominated solutions, demonstrating smooth end-to-end coverage of the objective space. Thus, they yield diverse parameter value pairs for the unknown transition parameters and offer decision-makers a detailed overview of the tradeoffs between the system requirements.

% Similarly, t
The Pareto front approximations produced for FPR2 and FPR3, albeit sparser, also cover the objective space satisfactorily, with the majority of the solutions clustered at the bottom left of the graph (where the total $D_{KL}$ is around 1). 
As expected, however, the range of $D_{KL}$ values per PK-informed property increases with solutions exceeding 3.5 (2.0) and 6 (2.5) for property R1 (R2) for FPR2 and FPR3, respectively. 
Increasing the number of unknown transition parameters in FPR instances FPR2 and FPR3 unavoidably expands the search space (Table~\ref{tab:systemVariants}) considerably and reshapes the landscape of the objective space. 
Consequently, \approach\ identifies solutions at the extremes of the objective space with very small $D_{KL}$ for one property and very large $D_{KL}$ for another. 

We visualise the distribution of PK-informed properties R1 (Fig.~\ref{fig_fpr_pf} middle) and R2 (Fig.~\ref{fig_fpr_pf} bottom) using \approach-derived Pareto front solutions against the prior knowledge provided by domain experts for these properties. 
Evidently, the \approach-produced distributions are good approximations of the target distributions.
The FPR1 solution produces distributions that accurately match the target distributions (total $D_{KL}$=0.08), while the FPR2 and FPR3 are close enough but with higher total $D_{KL}\!\!\approx$0.27 (given the extra 'degrees of freedom').
We obtained similar Pareto front approximations and sampled distributions for the PK-informed properties using \approach\ for the FX problem instances FX1, FX2 and FX3; due to space constraints, these results are available on the project webpage. 
These results corroborate our findings and demonstrate \approach's capabilities to extract combinations of unknown transition parameter values that closely approximate the properties distributions given by domain experts.

% \\ - Measure the KL divergence between the (ground truth) true distribution of PK-informed properties vs the approximated distribution from EPIK
% \\- Do this for the 4 cases (of increasing complexity in terms of number of properties and number of parameters prior values to be estimated)

\begin{figure}[t]
    \centering\includegraphics[width=\linewidth]{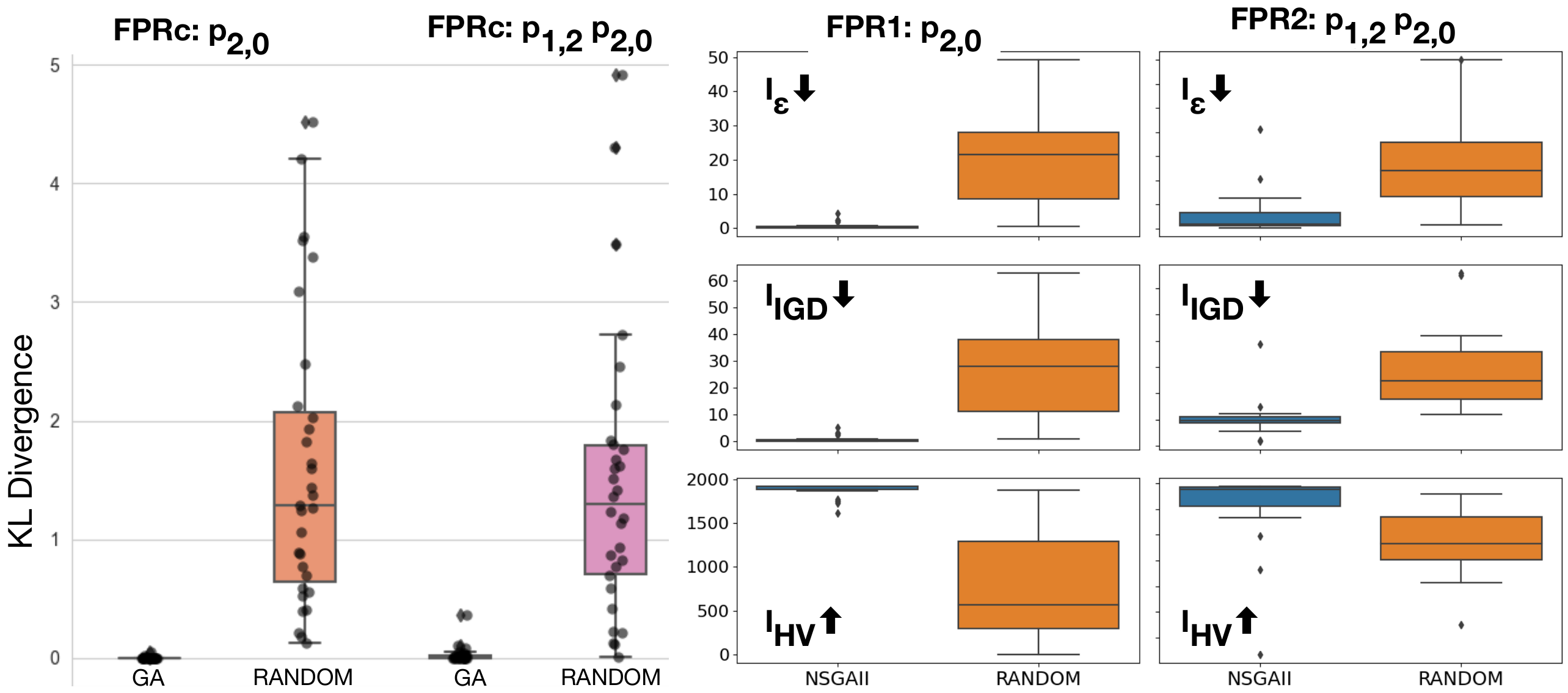}
    \vspace{-8mm}
    \Description{KL divergence between \approach\ solutions and Random search against ground truth for FPRc (left); and quality indicator boxplots for NSGAII and Random for FPR variants (right)}
    \caption{
    \changed{KL divergence between \approach\ solutions and Random search against ground truth for FPRc (left); and quality indicator boxplots for NSGAII and Random for FPR variants (right)}
    }
    % of \approach\ results for various conformance levels of FPR PK-informed properties}}
    \label{fig_fpr_rq2_RS}
    \vspace{-5mm}
\end{figure}

\begin{figure}[t]
    \centering\includegraphics[width=1\linewidth]{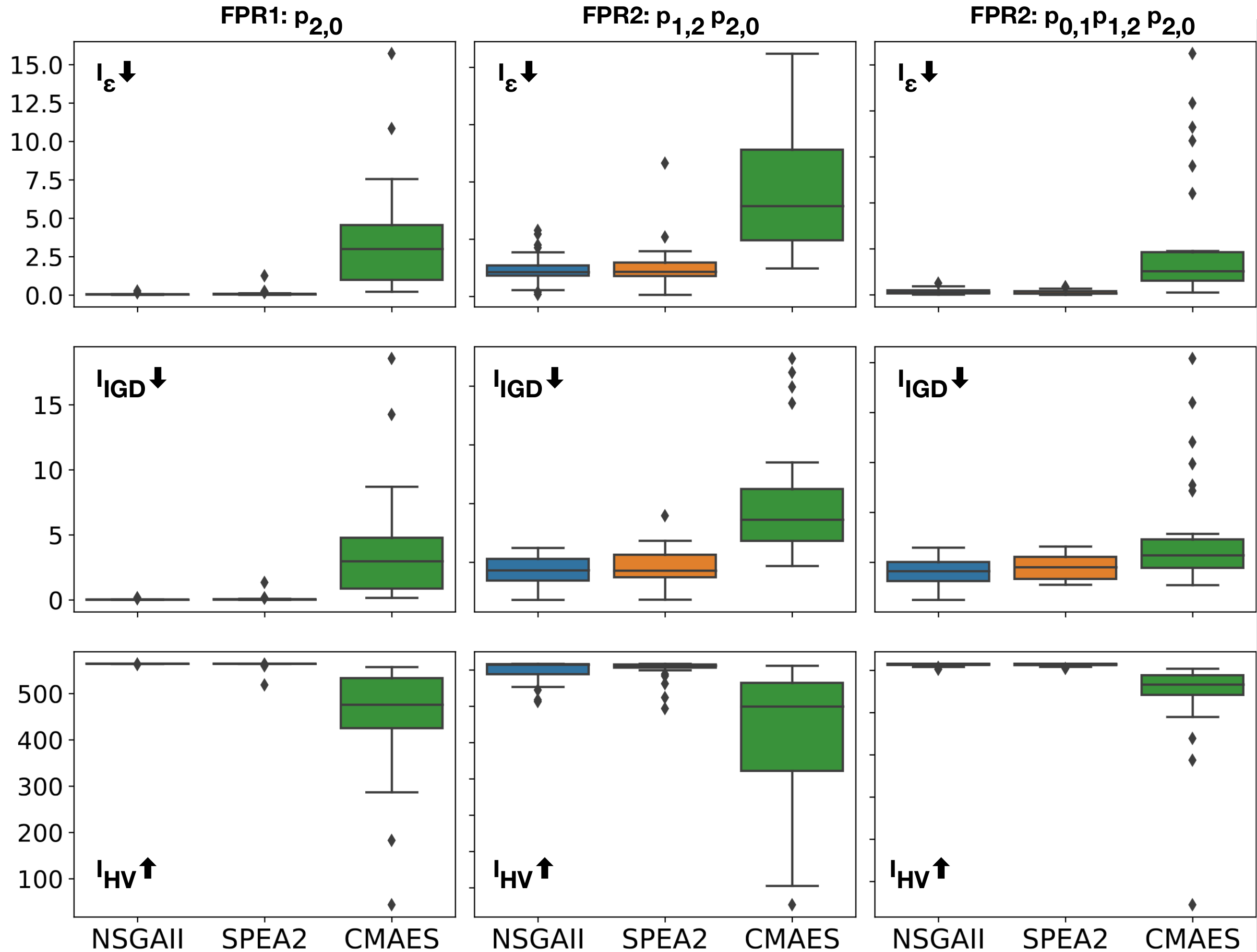}
    %\vspace{-5mm}
    % \caption*{\centering \small (a)}
    
    \vspace{1mm}
    
    % \label{fig_fpr_boxplots}
    % \end{figure}
    % \begin{figure}[t]
    \centering\includegraphics[width=1\linewidth]{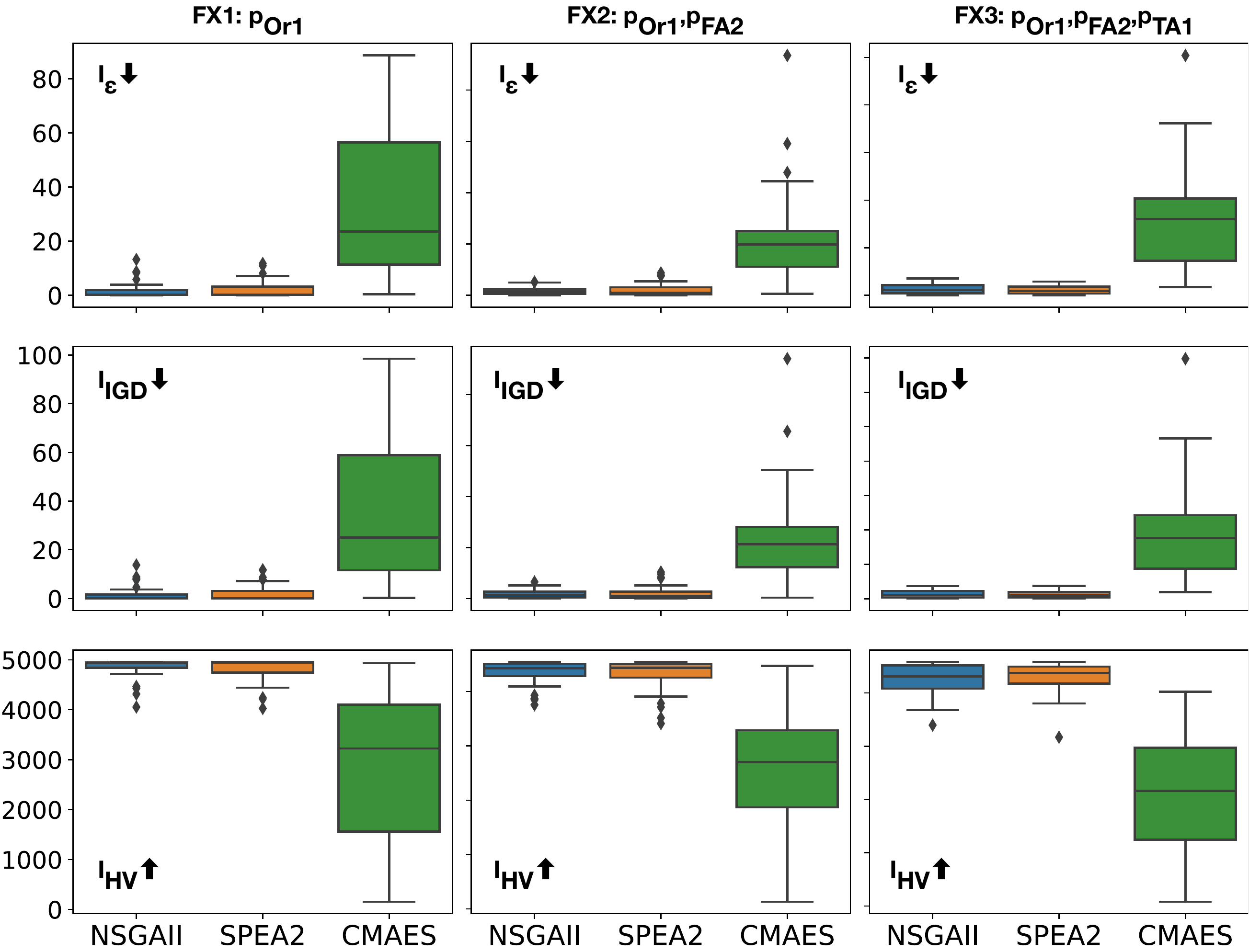}
    \vspace{-7mm}
    % \caption*{\centering \small (b)}
    \Description{Boxplots comparing \approach\ with multi-objective evolutionary algorithms for the FPR (top) and FX (bottom) systems}
    \caption{Boxplots comparing \approach\ with multi-objective evolutionary algorithms for the FPR (top) and FX (bottom) systems}
    \vspace{-6mm}
    \label{fig:boxplots}
\end{figure}

\vspace{1mm}
\noindent
\textbf{RQ2 (Effectiveness).}
\changed{First, Fig.~\ref{fig_fpr_rq2_RS} shows the KL result between \approach\ and Random search against ground truth for FPRc (left); and quality indicator boxplots for NSGAII and Random for FPR variants (right). 
These results show a clear gap between \approach-derived solutions and Random, signifying that the optimisation problem addressed by \approach\ is non-trivial, mandating the use of sophisticated evolutionary algorithms underpinning \approach.}
% We examine the impact of different multi-objective optimisation algorithms in EPIK’s performance. We study the quality of EPIK-derived model transition parameters using the established evolutionary algorithms NSGA-II [19], SPEA2 [77] and CMA-ES [41].

Next, we investigate \approach's effectiveness in eliciting admissible values for encoding the distributions of the unknown transition parameters using diverse multi-objective optimisation algorithms. 
% To this end, we instrumented \approach\ with the widely-used multi-objective volutionary algorithms NSGA-II~\cite{deb2002fast},  SPEA2~\cite{zitzler2001spea2} and CMA-ES~\cite{igel2007covariance}.
Fig.~\ref{fig:boxplots} depicts boxplots, over 30 independent runs, of the quality indicators $I_\epsilon$, $I_{IGD}$ and $I_{HV}$ for the various FPR and FX system instances when \approach\ is instrumented with the widely-used %multi-objective 
evolutionary algorithms NSGA-II~\cite{deb2002fast},  SPEA2~\cite{zitzler2001spea2} and CMA-ES~\cite{igel2007covariance}.
We observed that, despite \approach's flexibility, the CMA-ES-based \approach\ underperformed compared to the other \approach\ instances across all problem variants and for all quality indicators. 
The statistical analysis using Kruskal-Wallis yielded a statistically significant difference for all problem variants and quality indicator combinations (p-value $<$~0.0002), substantiating the performance difference that is evident in the boxplots. 
Through the post-hoc analysis, involving pairwise comparisons using Mann-Whitney and Cohen's d effect size, we established a statistically significant difference (p-value $<$ 1.205E-10) with a very large effect size (d$>$1.21) for all NSGA-II and SPEA2 comparisons against CMA-ES. 
In contrast, we found no statistically significant difference between \approach\ using NSGA-II or SPEA2; the p-value for all problem variants and indicators was [0.051, 0.79], exceeding the significance level~$\alpha=0.05$.

These findings clearly evidence that \approach\ employing NSGA-II and SPEA2 can produce Pareto front approximations that yield significantly better quality indicators than CMA-ES.
Furthermore, these results establish \approach's generality in identifying effective distributions for the unknown transition parameters by leveraging different multi-objective evolutionary algorithms.

\begin{figure}[t]
    \centering\includegraphics[width=\linewidth]{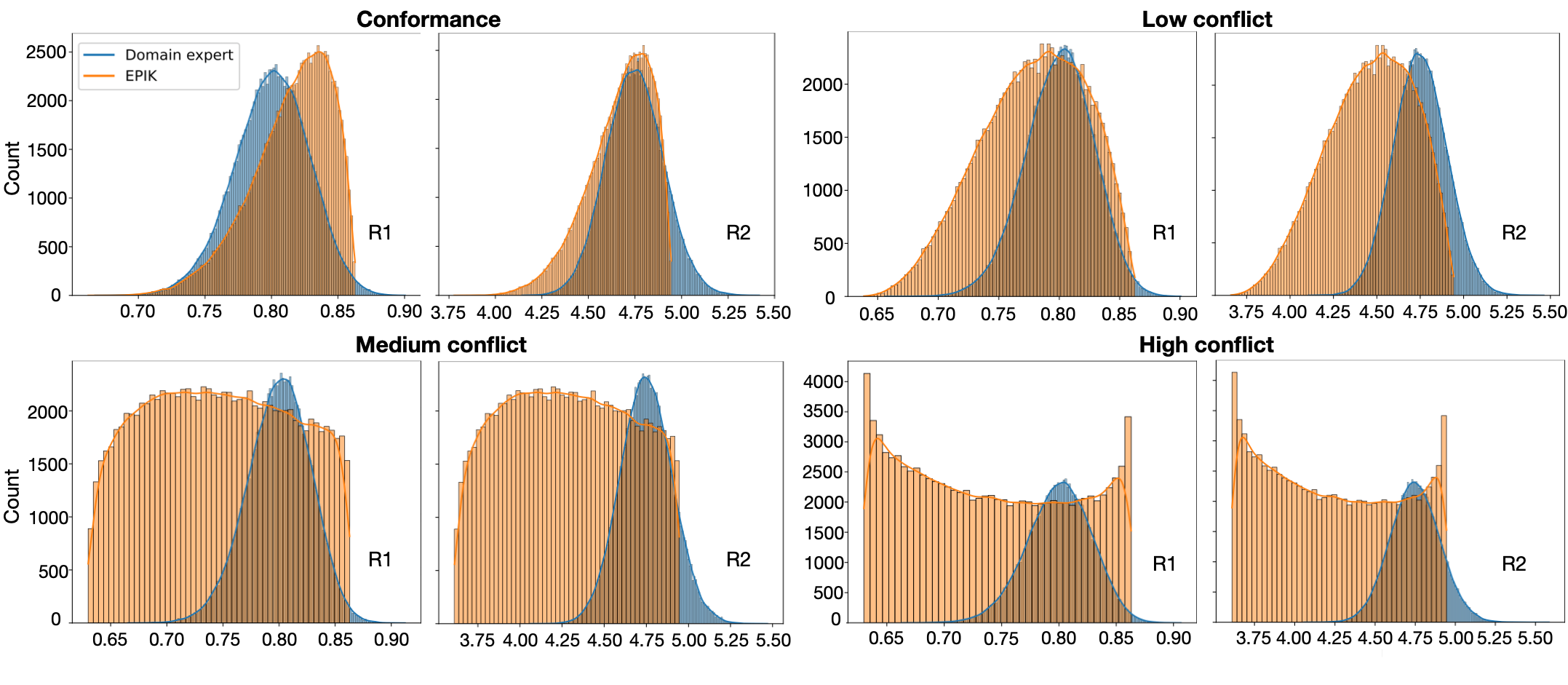}
    \vspace{-9mm}
    \Description{Distribution of \approach\ results for various conformance levels of FPR PK-informed properties}
    \caption{Distribution of \approach\ results for various conformance levels of FPR PK-informed properties}
    \label{fig_fpr_conformance_dist}
    \vspace{-7mm}
\end{figure}

% \begin{figure}[t]
% 	\centering\includegraphics[width=0.8\linewidth]{images/EPIK-RQ2b.pdf}
% 	\caption{Boxplots for conformance comparison on FPR}
% \label{fig_fpr_boxplots_conformance}
% \end{figure}

\vspace{1mm}
\noindent
\textbf{RQ3 (Conformance).}
We examine \approach's convergence when the PK-informed properties incorporate conflicting prior knowledge. 
This conflict can occur when domain experts have differing opinions about the PK-informed properties, leading to divergent accumulated knowledge. 
% thus, their accumulated knowledge exhibits divergence.  
A similar conflicting situation can occur when the same expert has strong knowledge about one PK-informed property but weak or contradictory knowledge about the other properties. 

We investigated this situation through four different conformance scenarios for the PK-informed properties of system FPR1.
Each scenario encodes a different conformance level, ranging from high conformance to high conflict. 
Fig.~\ref{fig_fpr_conformance_dist} shows the resulting distributions of the best solution with the lowest total $D_{KL}$ (given equal weight to objectives) found by the NSGA-II-based \approach. 
The solutions produced by \approach\ for the conformance and low-conflict scenarios yielded distributions that closely approximate the distributions of PK-informed properties from domain experts. 
Unavoidably, the higher the conflict level, the more divergent the distributions (especially for the medium and high conflict scenarios), indicating that such conflicting knowledge prevents \approach\ from finding solutions that further reduce the total $D_{KL}$ values. 

\begin{figure}[t]
    \vspace{-2mm}% \centering\includegraphics[width=0.8\linewidth]{images/KL_Evolution_median1.pdf}
    \centering\includegraphics[width=\linewidth]{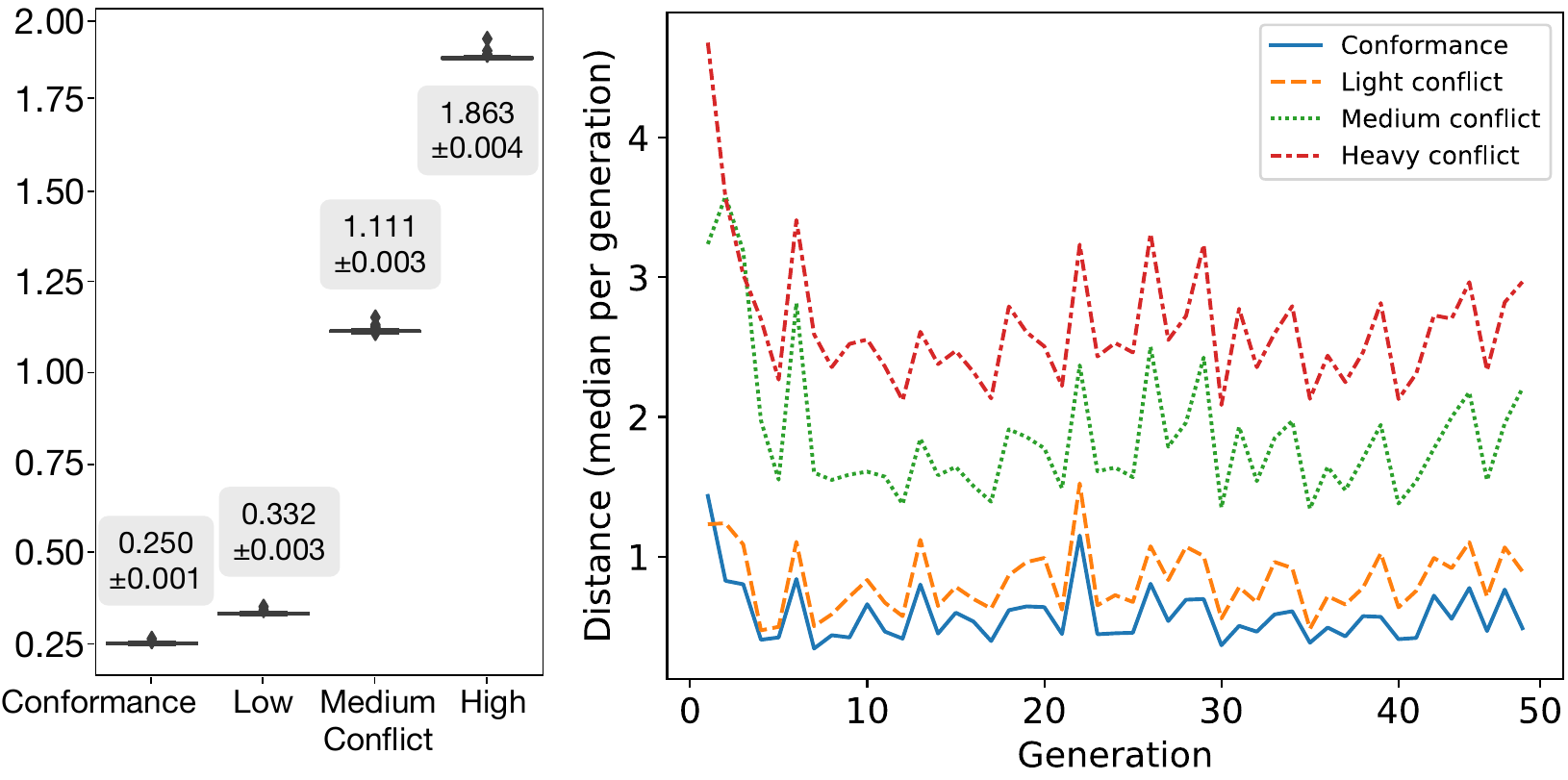}
    \vspace{-8mm}
    \Description{Boxplots for conformance comparison on FPR (left) and KL Evolution (median per generation) for a specific run and various conformance levels (right)}
    \caption{Boxplots for conformance comparison on FPR (left) and KL Evolution (median per generation) for a specific run and various conformance levels (right)}
    \label{fig_fpr_boxplots_conformance_KL_trace}
    \vspace{-6mm}
\end{figure}

The boxplots in Fig.~\ref{fig_fpr_boxplots_conformance_KL_trace} (left), generated over 30 independent runs, clearly show the distance between the total $D_{KL}$ values for the different conformance scenarios. 
Through the execution of the Kruskal-Wallis and Mann-Whitney inferential tests, we established statistically significant differences (p{\footnotesize$\;\ll\;$}0.05 and very large effect size) for all pairwise comparisons, including the conformance scenarios.
The narrow width of all boxplots confirms that the resulting gap is not due to randomness in \approach's generalisation or operation. 
Lastly, Fig.~\ref{fig_fpr_boxplots_conformance_KL_trace} (right) shows the median $D_{KL}$ evolution for each population per generation for a specific \approach\ run across all conformance scenarios. 
The evident fluctuation in the medium and high conflict scenarios illustrates the difficulty in finding stable Pareto front approximations that yield low $D_{KL}$ values.  
This observation is corroborated by the evident gap against the conformance and low conflict scenarios.
% is evident, demonstrate thatand which are comparable to the corresponding values derived from the conformance and low conflict scenarios. 

These results provide strong empirical evidence that \approach's ability to synthesise effective distributions for the unknown parameter values depends on the conformance level of the prior knowledge of the PK-informed properties. 
Identifying conflicting prior knowledge would be an interesting direction for future work to avoid wasting resources and to involve humans in the loop to resolve such conflicts.

% Devising effective methods to identify and resolve conflicts in the provided prior knowledge constitutes a key aspect of our future work. 

\begin{figure}[t]
    \centering\includegraphics[width=\linewidth]{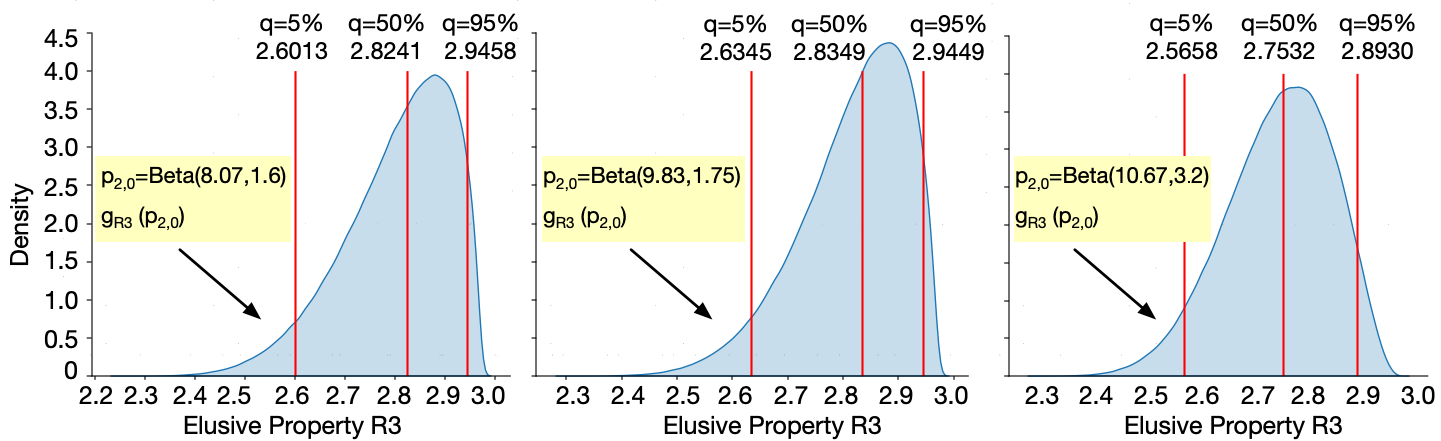}
    \vspace{-7mm}
    \Description{Elusive property R3 distribution for 3 FPR1 Pareto-optimal solutions from \approach's knowledge embedding stage}
    \caption{Elusive property R3 distribution for 3 FPR1 Pareto-optimal solutions from \approach's knowledge embedding stage}
    \vspace{-6mm}
    \label{fig_fpr_embedding}
\end{figure}

\vspace{1mm}
\noindent
\textbf{RQ4 (Knowledge Embedding).}
We employed the Pareto-optimal approximation solutions set synthesised during \approach's elicitation stage to perform knowledge embedding by solving the optimisation problem in Eq.~\eqref{eq_embed_moop_stage_2} and quantifying elusive property R3 (Table~\ref{tab:prestoReqs}).
Fig.~\ref{fig_fpr_embedding} shows the distributions of the verification results of elusive property R3 and the estimated quantiles $q=\{0.05, 0.5, 0.95\}$ for three FPR1 Pareto-optimal solutions 
$p_{2,0}\!\!=\!\!Beta(8.07,1.6)$, $p_{2,0}\!\!=\!\!Beta(9.83,1.75)$ and $p_{2,0}\!\!=\!\!Beta(10.67,3.16)$. 
The different Pareto-optimal solutions (signifying different $p_{2,0}$ instantiations) enable deriving elusive property distributions, which, albeit looking similar, yield different quantile values and density shapes.

Decision-makers can leverage this information to examine the shape of the elusive property's distribution.
Since no information existed for the unknown transition parameter values of the considered system, without \approach's embedding stage, this analysis could not have been performed at all or would involve a biased guess of the transition parameters, incurring the risk of inaccurate outcomes. 
\approach\ instead leverages the prior knowledge about the PK-informed properties to perform a well-reasoned analysis, enabling decision-makers also to select, as per Eq.~\eqref{eq_embed_moop_stage_2}, an appropriate solution that conforms to the system and elusive property semantics.
For instance, adopting a conservative approach for quantile $q=0.5$, and since the elusive property R3 quantifies the expected energy consumption for completing the fruit-picking process, decision-makers would select the middle solution $p_{2,0}\!\!=\!\!Beta(9.83,1.75)$ that yields the maximum median value.
The selected solution can then drive runtime quantitative verification~\cite{filieri_probabilistic_2013,zhao2023bayesian} and instrument self-adaptation~\cite{calinescu2017synthesis,filieri_supporting_2016}.

\vspace{1mm}
\noindent
\changed{
\textbf{\approach\ Discussion \& Guidelines.}
\\ 1) \textbf{Encoding summary statistics as distributions:}
When experts provide only a summary statistic (e.g., ``reliability is ~0.8"), \approach\ maps it to a canonical Beta/Gamma parameterisation (Eq.~\eqref{eq_prior_beta}). 
The point estimate serves as the ``best guess'' ($p_{i,j}^{(0)}$ or Gamma mean), while confidence is encoded via the sample size $n_i^{(0)}$. 
Setting $n_i^{(0)} \approx 1$ yields a near-uninformative prior, allowing subsequent Bayesian estimators~\cite{epifani_model_2009} to be driven primarily by runtime data; setting $n_i^{(0)}$ to historical observations (e.g., $n_i^{(0)}=200$ for 200 past missions) encodes a strong, concentrated belief. 
This supports a continuous spectrum from uninformative to highly informative priors without requiring experts to specify a distribution shape by hand.
%
% \\
% When domain experts can only provide a summary statistic rather than a full distribution (e.g., ``reliability is ~0.8"), \approach\ still accommodates this directly via the canonical Beta/Gamma parameterisation (Eq.~\eqref{eq_prior_beta}). 
% The expert's point estimate is set as the ``best guess'' parameter ($p_{i,j}^{(0)}$ for probabilities, the mean for Gamma-distributed rewards), while the confidence behind the estimate is encoded through the sample-size parameter $n_i^{(0)}$. 
% Setting $n_i^{(0)} \approx 1$ yields a near-uninformative prior centred on the stated value, letting subsequent Bayesian estimators~\cite{epifani_model_2009} driven predominantly by runtime data; setting $n_i^{(0)}$ to the number of underlying historical observations (e.g., $n_i^{(0)}=200$ for 200 past missions) encodes a stronger, more concentrated belief. 
% \approach\ thus supports a continuous spectrum from uninformative to highly informative priors from partial expert knowledge, without requiring experts to specify a distribution shape by hand.
%
\\ 2) \textbf{Conflict detection}:
% Because \approach's elicitation stage (Section~\ref{sec_knowledge_elicitation}) minimises the KL divergence between expert-supplied and induced PK-informed property distributions (Eq.~\eqref{alg_eliciting_opt}), the resulting Pareto front is itself a diagnostic for conflicting prior knowledge. 
% \approach\ adopters can monitor two indicators, as demonstrated in RQ3: (i) the magnitude of total $D_{KL}$ achieved by the best Pareto-optimal solution(s); solutions clustering only at large $D_{KL}$ (rather than near 0) indicate that no candidate parameter distribution can simultaneously satisfy all specified properties; and (ii) generational stability of the median $D_{KL}$ (Fig.~\ref{fig_fpr_boxplots_conformance_KL_trace}, right); sustained oscillation rather than smooth convergence signals contradictory rather than merely difficult objectives. Either symptom should prompt to revisit the elicited PK-informed properties with domain experts (e.g., reconciling disagreements or adjusting $n_i^{(0)}$ confidence values) before trusting the resulting priors.
Because \approach's elicitation stage (Section~\ref{sec_knowledge_elicitation}) minimises KL divergence (Eq.~\eqref{eq_obj}), the Pareto front diagnoses conflicting prior knowledge. 
Adopters can monitor two indicators (RQ3): (i) the magnitude of total $D_{KL}$ achieved by Pareto-optimal solutions, where values clustering far from 0 indicate that no candidate distribution can simultaneously satisfy all properties; and (ii) the generational stability of the median $D_{KL}$ (Fig.~\ref{fig_fpr_boxplots_conformance_KL_trace}, right), where sustained oscillation rather than smooth convergence signals contradictory objectives. 
Either symptom prompts users to reconcile the PK-informed properties with experts before trusting the resulting priors.
\\ 3) \textbf{Managing Non-Identifiability and Epistemic Ambiguity}:
% In inverse problems, multiple distinct transition parameter distributions can yield the same observable system-level property distributions. 
% \approach\ treats this non-identifiability as a feature of the problem space.
% The synthesised Pareto set ($PS_{s1}$) must not be interpreted as a single ``correct" parameter recovery and a single $\theta \in PS_{s1}$ should not be read as the recovered ground truth. 
% Instead, the Pareto set represents the family of admissible priors consistent with the formal model structure and the available domain knowledge. 
% % its spread quantifies residual epistemic ambiguity. 
% % Instead, it represents a family of admissible parameter priors constrained by the formal model structure and the domain knowledge. 
% The shape and spread of the Pareto set reflect the epistemic ambiguity inherent in the system mapping. 
% Decision-makers should evaluate multiple solutions along this front to understand how parameter variation impacts unobserved system behaviours.
In inverse problems, multiple distinct transition parameter distributions can yield identical system-level property distributions. 
\approach\ treats this non-identifiability as a feature: the synthesised Pareto set ($PS_{s1}$) is not a single ``correct'' parameter recovery, nor is any $\theta \in PS_{s1}$ the absolute ground truth. 
The Pareto set represents the family of admissible priors consistent with the formal model and domain knowledge.
\\ 4) \textbf{Conditional Soundness of Elusive Properties}:
% By definition, elusive properties (e.g., rare safety hazards or expensive metrics) cannot be validated directly at design time due to a complete lack of prior data.  
% We emphasise that \approach's soundness in verifying elusive properties is strictly conditional. 
% Concretely, the correctness of the elusive property estimates relies on the assumption that the PK-informed properties are accurately specified and that the underlying formal model structure is correct.
% Accordingly, if the expert-provided priors for the $Y$ PK-informed properties are biased, the resulting transition parameters—and consequently, the elusive property evaluations—will also be biased. 
% This is an inescapable epistemic limitation shared by all Bayesian elicitation techniques
Elusive properties cannot be validated directly at design time due to missing prior data. 
We stress that \approach's soundness in verifying these properties is strictly conditional: it assumes that PK-informed properties are accurate and the underlying model structure is correct. 
If expert-provided priors are biased, the resulting parameters, and consequently, the elusive property evaluations, will also be biased. 
This is an unavoidable epistemic limitation of all Bayesian elicitation techniques.
\\ 5) \textbf{Performance \& Scalability Trade-offs.} 
% To optimise \approach's execution overheads, practitioners should configure \approach\ based on the complexity and formal structure of the underlying Markov models.
% More specifically, the offline algebraic formulae extraction incurs a high upfront cost (seconds to $\sim 8$ minutes), but makes online evaluation extremely fast (milliseconds per sample).  
% Thus, this step is recommended for standard DTMC/CTMC models where closed-form rational functions can be symbolically computed.  
% When, however, the system model's size or property complexity prevents symbolic algebraic expression extraction, \approach\ adopters should resort to iterative PMC fallback which, albeit yielding correct results, incurs high continuous cost due to the model checker invocation in-the-loop for every candidate solution.
Practitioners should configure \approach\ based on the model's complexity. 
Offline algebraic formula extraction has a high upfront cost (seconds to $\sim 8$ minutes for the systems evaluated) but enables millisecond-level evaluations; this is recommended for DTMC/CTMC models where closed-form rational functions can be symbolically computed. 
If model scale or property complexity prevents symbolic extraction, adopters should fall back to iterative PMC, which guarantees correct results but incurs high continuous overhead by running PMC in-the-loop.
}

\vspace{-5mm}
\subsection{Threats to Validity}
\label{sec_ttv}
% %\vspace{-1mm}
\noindent
We limit \textbf{construct validity} threats that may occur due to simplifications and assumptions in the evaluation by using formal models and properties of systems from real-world case studies from the literature
(e.g., FPR~\cite{fang2022presto}, FX~\cite{gerasimou_search-based_2015}).
% taken from the literature. For example, the FX system, the model and properties were developed in close collaboration with a foreign exchange domain expert in \cite{gerasimou_search-based_2015}. 
Also, \approach\ employs KL for comparing the candidate and ground truth distributions, thus leveraging its asymmetric property and optimising the candidate distribution to the ground truth.
Experimenting with other distance metrics (e.g., Jensen–Shannon divergence) would help in further validating \approach.
% Due to the simplification of using numerical solutions, settling for approximate solutions rather than global optima reduces confidence in the identified optimal parameters. To mitigate, we run multiple optimisation trials for each experiment and RQ3 is dedicated to statistically investigate the effectiveness of various optimisation approaches in EPIK. While different metrics exist for comparing distributions, the threat of using KL divergence rather than others (e.g., Jensen–Shannon divergence) is mitigated given its popularity and asymmetric property that suits our formulated problem (i.e., optimising the distribution to the ground truth). That said, more experiments using multiple metrics in addition to KL divergence would further validate the method.

We reduce \textbf{internal validity} threats that may introduce bias in establishing cause-effect relationships in our experiments by reporting results over 30 independent runs per experiment, and using statistical tests to check for statistical significance. 
Also, since \approach\ depends on the prior knowledge from experts for PK-informed properties, we examine (RQ3) the \approach's ability to cope with inaccurate and/or conflicting prior knowledge. 
Finally, we enable replication by making all experimental results publicly available on our project webpage.

% may correspond to bias in establishing cause-effect relationships in our experiments. For instance, domain experts providing PK may introduce potential bias. To mitigate, we conduct RQ2 to examine how effective EPIK can cope with inaccurate and potentially conflicting PK. We also reported results over 30 repeated runs of each experiment, and employed statistical tests to check for significance in the achieved results. Finally, we enable replication by making all experimental results publicly available on our project webpage.

% \textbf{External Validity}
We limit \textbf{external validity} threats that may reduce the \approach's generalisability by leveraging knowledge from experts to build the PK-informed properties. %assuming that these experts will possess such knowledge. 
The strength of prior knowledge can be easily encoded into the PK-informed properties as well as the incorporation of prior knowledge from multiple experts (i.e., the cumulative knowledge encoded in the parameters of the corresponding distribution, e.g., Gamma) for the PK-informed property. 
Although we used established case studies, doing more experiments in domains and applications with different characteristics than those from our evaluation would further validate \approach's applicability and scalability.

\vspace{-4mm}
\section{Related Work}
\label{sec_related_work}

\textbf{Applications of quantitative verification in software engineering}. 
PMC-based quantitative verification has been widely applied in various domains, including self-adaptive systems \cite{calinescu_self_adaptive_2012,calinescu2017engineering} and, more recently, autonomous systems \cite{kwiatkowska_probabilistic_2022,vazquez2025adaptive}. 
Accordingly, there has been a significant focus of research and publications within the software engineering community. 
The QoSMOS framework~\cite{calinescu_dynamic_2011} is an illustrative example for developing adaptive service-based systems, using PRISM to compute quality-of-service properties. 
Similarly, the application in~\cite{filieri_probabilistic_2013} is a typical
web-based self-adaptive system that comprises an HTTP Proxy server, a web server and an application server. 
A low-power wireless bus system is studied in~\cite{filieri_supporting_2016} to be efficiently verified at run time as soon as changes occur, while dynamic power management systems are also investigated~\cite{calinescu2017synthesis,gerasimou_search-based_2015}.
Interested readers can find further examples of PMC applications in software engineering for self-adaptive systems in~\cite{exemplars}.

For autonomous systems, PMC-based quantitative verification has been applied to a range of scenarios, including spacecraft reconfiguration~\cite{nardone2016probabilistic}, motion planning~\cite{lahijanian2011temporal,lahijanian2015formal}, controller synthesis for unmanned underwater vehicles and unmanned aerial vehicles~\cite{giaquinta2018strategy,getir2025specification,zhao_towards_2019,calinescu_self_adaptive_2015}, and safety/reliability assurance in harsh environments~\cite{calinescu2017engineering,zhao_probabilistic_2019,dong2022dependability,zhao2023bayesian,gerasimou2021evolutionary}. 
Additional applications include task allocation and planning for mobile robots~\cite{lacerda2019probabilistic,getir2025specification,vazquez2026mind} and their battery charge scheduling~\cite{zhao_towards_2019,tomy2020battery}.
While this is not an exhaustive list of practical applications for quantitative verification, all of these scenarios can benefit from  \approach\ when integrated with Bayesian learning methods to develop more accurate probabilistic models.

\textbf{Parameter estimation in quantitative verification.}
Despite progress in quantitative verification~\cite{kwiatkowska_probabilistic_2022}, a key challenge persists: quantitative verification assumes that the models accurately reflect real software behaviour. 
This is often true for model structures, but transition probabilities/rates are harder to estimate correctly~\cite{calinescu_self_adaptive_2012}. 
Relying on point estimates provided by domain experts, inferred through model fitting~\cite{su2013asymptotic} or updated at runtime~\cite{filieri_run_time_2011} incurs unquantified estimation errors, which are propagated and compounded in later verification steps in ways that are unknown but likely to be significant~\cite{calinescu_self_adaptive_2012}. 
Although recent research~\cite{calinescu_formal_2016} synthesises bounds for unknown transition parameters via confidence intervals, it considers solely operational data, disregarding any prior human knowledge. 
This limitation motivates the use of Bayesian estimators~\cite{filieri_run_time_2011}. 
% This limitation motivates the exploration of using \textit{Bayesian estimators} to embed PK, which will be discussed next.

\textbf{Bayesian learning for runtime quantitative verification.} 
The KAMI framework~\cite{epifani_model_2009} introduced Bayesian learning to estimate the transition probabilities of DTMCs. 
The framework was later retrofitted for CTMCs~\cite{filieri_formal_2012} and extended with ageing factors to capture time-varying transition probabilities~\cite{calinescu_adaptive_2014} and with a lightweight adaptive filter to reduce the noise~\cite{filieri_lightweight_2015}.
Robust Bayesian estimators yielding interval estimates for DTMCs and CTMCs have been introduced in~\cite{zhao_probabilistic_2019,zhao2023bayesian}, while~\cite{epifani_change-point_2010,zhao_interval_2020} devised Bayesian-based change-point detectors to determine change points during the system operation. 
\approach\ is unique as it facilitates the elicitation and embedding of prior knowledge required by these techniques.
% Collectively, this yields a comprehensive framework for Bayesian learning in runtime PMC.

\textbf{Domain knowledge elicitation and embedding.}
This topic is studied across software engineering~\cite{o2006uncertain,WRIGHT198713}, including requirements engineering~\cite{nuseibeh2000requirements,hadar2014role}, where knowledge can help clarify the needs of stakeholders and the system boundaries so as to improve software design~\cite{adelson1986model,sonnentag1998expertise}.
Likewise, in software testing, domain knowledge aids in test case prioritisation and selection~\cite{jin2019finexpert}, while leveraging prior knowledge in safety-critical systems testing can significantly accelerate the testing process, even when the PK is partial and vague~\cite{zhao_assessing_2020}. 
Incorporating domain prior knowledge is also gaining attention in machine learning testing and verification~\cite{braiek2020testing, stewart2017label,xie2021survey,kerrigan2021survey}, 
backdoor attack/defence~\cite{huang_embedding_2022}, 
and explainability~\cite{heaton2023explainable}.

To the best of our knowledge, only~\cite{oghabi2011verification,zervoudakis2013cascading,calinescu_efficient_2021} combine domain knowledge with PMC-based quantitative verification. 
The Web Ontology Language is used to describe service behaviours as domain knowledge and then generate stochastic formal models~\cite{oghabi2011verification}. 
% Oghabi et al.\cite{oghabi2011verification} use the Web Ontology Language to describe service behaviours as domain knowledge and then generate stochastic formal models. 
Similarly, a YAML-based domain-specific language is used to express system specifications that can be compiled into PCTL properties and verified by a model checker~\cite{zervoudakis2013cascading}.
% Zervoudakis et al.~\cite{zervoudakis2013cascading} devise a YAML-based domain-specific language to express system specifications that can be compiled into PCTL properties and verified by a model checker.
Domain-specific modelling patterns in~\cite{calinescu_efficient_2021} and fragmentation techniques~\cite{fang2023fast} speed up the analysis of parametric Markov chains, %that model software system behaviours, 
enabling the extraction of closed-form expressions for properties of such patterns. 
% The work \cite{calinescu_efficient_2021} speeds up the analysis of parametric Markov chains that model software behaviours by exploiting domain-specific modelling patterns (as PK), and then pre-computes closed-form expressions for properties of such patterns, and uses these expressions in the analysis of whole-system models. 
This research devises formal models and properties exploiting other explicit formalisms. 
\approach\ is unique and innovative as it is natively designed to elicit and embed prior knowledge into model parameters without relying on external semantics or domain-specific languages.

\vspace{-4mm}
\section{Conclusion \& Future Work}
\label{sec_conclusion}
% %\vspace{-2mm}
% We introduced \approach, the first work for eliciting and embedding prior knowledge in Bayesian learning for quantitative verification. 
% \approach\ leverages expertise of domain experts in the form of prior knowledge about observable system-level properties. 
% \approach\ uses this knowledge to formulate a two-fold optimisation problem, enabling the extraction of Pareto-optimal approximation sets, encoding distributions of unknown model transition parameters, and whose corresponding Pareto fronts yield system-level property distributions that closely approximate those provided by domain experts. 
% Decision-makers can leverage the derived Pareto fronts during \approach's knowledge embedding step to quantify elusive system-level properties which are novel or rare, or those for which gathering information is risky or expensive.
% Through a comprehensive experimental evaluation using multiple variants of real-world case studies and diverse \approach\ instantiations, we demonstrated the effectiveness, flexibility and generality of our approach. 
\changed{
We introduced \approach, the first work for eliciting and embedding prior knowledge in Bayesian learning for quantitative verification. \approach\ leverages expert knowledge of observable system-level properties to formulate a two-fold optimization problem, extracting Pareto-optimal approximation sets of unknown transition parameters whose Pareto fronts closely match expert expectations. Decision-makers can use these fronts to quantify elusive (rare, novel, or expensive) system properties.}
Through a comprehensive experimental evaluation using multiple variants of real-world case studies and diverse \approach\ instantiations, we demonstrated the effectiveness, flexibility and generality of our approach. 
% Finally, comprehensive evaluation across multiple real-world case studies and diverse instantiations demonstrates the effectiveness, flexibility, and generality of our approach.
%
Our future work includes 
(1) enabling knowledge elicitation and embedding for interval Bayesian verification~\cite{zhao2023bayesian};
(2) investigating parallelisation methods to improve \approach's scalability;
(3) evaluating \approach\ to other case studies and scenarios.

\textbf{Data Availability}. The open-source EPIK replication package is available at 
\url{https://doi.org/10.5281/zenodo.19337648}.
% \url{https://anonymous.4open.science/r/EPIK-BAD4}. 

\balance

\bibliographystyle{ACM-Reference-Format}
\bibliography{ref}

\end{document}